\documentclass[preprint,3p,onecolumn,final]{elsarticle}

\usepackage{amsmath,amssymb}
\usepackage{lineno,hyperref}

\journal{Nuclear Physics B}

\begin{document}
	
\begin{frontmatter}
\title{\textbf{Transverse momentum and rapidity distributions of top quarks in pair and single top quark production within the non-commutative standard model at the LHC}}
\author[a]{Mohamed El Arebi Gadja\texorpdfstring{\corref{cor}}}
\cortext[cor]{Corresponding author.}
\ead{gadja.mohamed@univ-ouargla.dz}
\author[b]{Yazid Delenda}
\ead{yazid.delenda@univ-batna.dz}
\author[a]{Lamine Khodja}
\ead{khodja.lamine@univ-ouargla.dz}

\affiliation[a]{organization={Laboratoire de Rayonnement et Plasmas et Physique de Surfaces, Département de Physique, Faculté des Mathématiques et des Sciences de la Matière, Université Kasdi Merbah Ouargla},
			city={Ouargla},
			postcode={30000},
			country={Algeria}}
\affiliation[b]{organization={Laboratoire de Physique des Rayonnements et de leurs Interactions avec la Matière, Département de Physique, Faculté des Sciences de la Matière, Université de Batna-1},
			city={Batna},
			postcode={05000},
			country={Algeria}}
\begin{abstract}
This paper presents a comprehensive study of top quark phenomenology within the Non-Commutative Standard Model. We calculate non-commutative corrections to the squared amplitudes for top quark pair production, as well as for single top quark production via the $t$-channel and $tW$-channel, while noting that the $s$-channel remains unaffected by non-commutative geometry. Utilizing \texttt{MadGraph5\_aMC@NLO}, we determine total cross-sections at various center-of-mass energies and examine differential distributions in transverse momentum and rapidity at leading order in both the strong coupling $\alpha_S$ and the non-commutative parameter $\Theta$. For single top production in the $t$-channel, a matching technique is employed to extend the validity of the distribution to low \(p_t^{\rm{top}}\) values through resummation. We also compare the non-commutative corrections with higher-order QCD corrections obtained at NLO using \texttt{MCFM} and at NNLO from existing literature. Our findings reveal significant deviations arising from non-commutative geometry at high energies, providing insights into potential new physics at energy scales accessible by current and future colliders.
\end{abstract}

\end{frontmatter}

\section{Introduction}
\label{introduction}
	
The study of the top quark is of paramount importance in high-energy physics, as it provides a critical testing ground for the Standard Model (SM) and has the potential to reveal physics beyond the SM. The significance of the top quark stems from its large mass, which suggests that it may play a fundamental role in electroweak symmetry breaking. This has profound implications for the physics program at the Large Hadron Collider (LHC). Top quarks and their antiparticles can be produced at the LHC predominantly through two mechanisms: quark-antiquark annihilation, accounting for approximately 10\% of the production, and gluon-gluon fusion, which dominates with about 90\%. This is in contrast to the Tevatron, where quark-antiquark annihilation is the primary production mechanism, responsible for 90\% of the top quark events. Although top quark production is relatively rare due to its large mass, the high luminosity of the LHC provides a sufficient number of events to enable detailed studies of its properties and interactions \cite{Khachatryan_2015,Khachatryan_2016,Aad_2016a,Aad_2016b}.
	
Research on top quarks at the LHC is crucial for deepening our understanding of the SM and exploring potential new physics. Extensive studies have already been conducted, such as the computation of next-to-leading order (NLO) QCD corrections \cite{NASON1988607,BEENAKKER1991507,MANGANO1992295,PhysRevD.40.54} and photoelectric corrections \cite{beenakker1994electroweak,kuhn2007weak}. The primary production mechanism for top quarks involves the strong force, leading to the formation of top-antitop pairs. However, single top quarks can also be produced through weak interactions via three main processes: $s$-channel, $t$-channel, and $tW$ associated production. The $t$-channel process, represented at leading order by $q + b \rightarrow q' + t$, was first observed at the Tevatron \cite{abazov2009observation,CDF:2009itk} and accounts for approximately 20\% of top-quark production at the LHC \cite{campbell2021single}. The $tW$ associated production, represented by $bg \rightarrow t + W$, features a final state with a top quark and a $W$ boson \cite{ott2012search}.
	
Despite occurring through weak interactions, $t$-channel single top production has a significant rate due to its favorable kinematics compared to pair-production processes. This channel is instrumental in probing the intrinsic properties of the top quark, such as its mass \cite{CMS:2017mpr} and polarization \cite{khachatryan2016measurement}. Additionally, it facilitates detailed tests of the SM at the differential level \cite{aaboud2017fiducial,sirunyan2020measurement} and constrains anomalous $Wtb$ couplings \cite{aad2016search,CMS:2016uzc,ATLAS:2017ygi}. Furthermore, it provides insights into the components of the production mechanism, including the parton distribution function (PDF) of the bottom quark and the CKM matrix element $V_{tb}$, which has been measured at both the Tevatron \cite{CDF:2015gsg} and the LHC \cite{chatrchyan2011measurement,CMS:2014mgj,sirunyan2020measurement}.
	
This work focuses on the production processes of both pair and single top quarks at the LHC within the framework of the non-commutative Standard Model (NC-SM). We specifically analyze two key observables used to extract the properties of the top quark: the transverse-momentum distribution of the top quark in pair and single top events, and the rapidity distributions of $t\bar{t}$ pairs. These observables have been measured by the ATLAS and CMS experiments in proton-proton collisions at a center-of-mass energy of $\sqrt{s}=7$ TeV \cite{daubie2013measurement,ATLAS:2015dbj}.
	
The unique features of non-commutative space-time quantum field theory (see ref \cite{douglas2001noncommutative}), particularly the NC-SM (for a comprehensive description and review of the NC-SM, see \cite{madore2000gauge,jurco2000enveloping,jurvco2001nonabelian,Calmet:2001na}), motivate this study. Prior research has investigated various aspects of the top quark within the NC-SM, including its decays \cite{Mahajan:2003ze,MohammadiNajafabadi:2008zlg,MohammadiNajafabadi:2006iu}, forward-backward asymmetry \cite{Fisli:2020vzt}, and azimuthal polarization \cite{Rezaei:2018mlk}. In this context, non-commutative space-time is constructed by replacing the ordinary commutation relation between coordinates with the following commutator for the Hermitian operators $\hat{x}$ \cite{Filk:1996dm}:
\begin{equation}
	[\hat{x}_\mu,\hat{x}_\nu]=i\Theta_{\mu\nu}\,,
\end{equation}
where $\Theta_{\mu\nu}$ is a constant, antisymmetric matrix that determines the non-commutativity of space-time. This matrix introduces a fundamental length scale below which space-time coordinates become non-commutative, and it can also be interpreted as a background field that dictates the relative orientation of different space-time directions.
	
To develop quantum field theories in non-commutative space-time, we employ the star product, defined by \cite{douglas2001noncommutative,riad2000noncommutative}:
\begin{align}
\psi(x)\star\phi(x)=\psi(x)\exp\left[\frac{i}{2}\,\overleftarrow{\partial}_\mu\,\Theta^{\mu\nu}\,\overrightarrow{\partial}_\nu\right]\phi(x)\,.
\end{align}
	
To preserve gauge symmetry in gauge theories defined on non-commutative space-time, we use Seiberg-Witten maps. These maps relate non-commutative gauge fields to ordinary fields in commutative theory via a power series expansion in $\Theta$, as follows \cite{seiberg1999string,melic2005standard,uelker2008seiberg,ulker2012all}:
\begin{subequations}
\begin{align}
		\hat{\psi}(x,\Theta) &=\psi(x)+\Theta\,\psi^{(1)}[V]+\mathcal{O}(\Theta^2)\,,\\
		\hat{V}_\mu(x,\Theta)&=V_\mu+\Theta\,V_\mu^{(1)}[V]+\mathcal{O}(\Theta^2)\,,
\end{align}
\end{subequations}
where $\psi$ is the ordinary spinor field, and $V_\mu$ is the gauge field. To leading order in the $\Theta$ parameter \cite{Jurco:2001rq,uelker2008seiberg}:
\begin{subequations}
\begin{align}
	\Theta\,\psi^{(1)}[V]&=-\frac{1}{2}\,\Theta^{\mu\nu}\,V_\mu\,\partial_\nu\psi+\frac{i}{4}\,\Theta^{\mu\nu}\,V_\mu\,V_\nu\,\psi\,,\\
	\Theta\, V^{(1)}_\mu[V]&=-\frac{1}{4}\Theta^{\alpha\beta}\{V_\alpha,\partial_\beta V_\mu+F_{\beta\mu}\}\,,
\end{align}
\end{subequations}
where $F_{\beta\mu}=(i/g_s)[D_\mu,D_\nu]$ is the field strength tensor, and the curly brackets denote the anti-commutator.
	
The structure of this paper is as follows: In Section 2, we present the Standard Model within the non-commutative geometry formalism and calculate the leading-order cross-section for top-antitop pair production in $pp$ collisions at the LHC. In Section 3, we compute the leading-order cross-sections for single top quark production in the $t$-channel and $tW$ associated production channel in $pp$ collisions at the LHC. In Section 4, we extract the transverse momentum distribution of top quarks in both pair and single top production, as well as the rapidity distribution of top-antitop pairs. For these calculations, we use \texttt{MadGraph} \cite{Maltoni:2002qb, Alwall:2014hca} and \texttt{MadAnalysis 5} \cite{Conte:2012fm} to perform the convolution with parton distribution functions (PDFs) and to apply experimental cuts. Finally, in the last section, we present our conclusions.
		
\section{Top-Antitop Quark Production in the NC-SM}
\label{sec:ttbar_NCSM}
		
\begin{figure}[ht]
\centering
\includegraphics[width=0.23\textwidth]{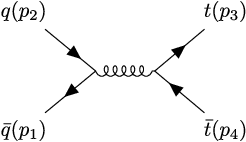}
\includegraphics[width=0.23\textwidth]{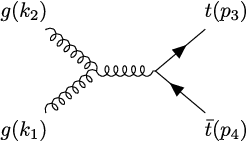}
\includegraphics[width=0.23\textwidth]{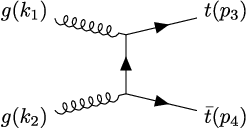}
\includegraphics[width=0.23\textwidth]{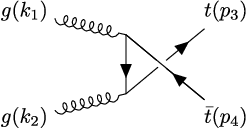}
\caption{Feynman diagrams contributing to the subprocess $pp \rightarrow \text{top} + \text{antitop}$.}
\label{fig1}
\end{figure}

In the Standard Model (SM), the production of top-antitop quark pairs in proton-proton (pp) collisions involves the Feynman diagrams shown in Fig.~\ref{fig1}. These diagrams remain the same in the non-commutative Standard Model (NC-SM) framework, but with modifications to the vertices due to the non-commutative geometry.
		
\begin{figure}[ht]
\centering
\includegraphics[width=0.23\textwidth]{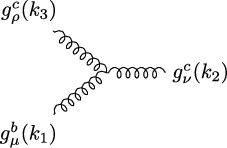}
\hspace{1cm}
\includegraphics[width=0.23\textwidth]{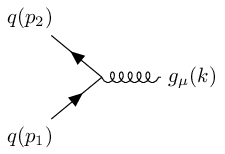}
\caption{Examples of strong interaction vertices modified by non-commutative geometry \cite{melic2005standard}.}
\label{fig01}
\end{figure}

In the NC-SM, non-commutative geometry (NCG) introduces corrections at first order in the non-commutative parameter $\Theta$, which affects only the triple-gluon vertex and the gluon-quark-quark vertex, as illustrated in Fig.~\ref{fig01}.
		
In these figures, $g_s$ denotes the strong coupling constant, $f^{abc}$ and $d^{abc}$ represent the SU(3) structure constants and $d$-symbols, respectively, and $T^a$ are the generators of the SU(3) group. The momenta of the gluons are denoted by $k_i$, and $\Theta^{\mu\nu\rho} = \Theta^{\mu\nu} \gamma^\rho + \Theta^{\nu\rho} \gamma^\mu + \Theta^{rho\mu} \gamma^\nu$ encapsulates the non-commutative effects.
		
The amplitudes for the subprocesses $q\bar{q} \rightarrow t\bar{t}$ and $gg \rightarrow t\bar{t}$ in the NC-SM are given by:
\begin{align*}
			i\mathcal{M}_{q\bar{q} \rightarrow t \bar{t}}^{\text{NC}} 
			&= i\frac{g_s^2}{\ell_1^2} \bar{v}(p_2)\mathcal{B}^\mu u(p_1) \bar{u}(p_3)\mathcal{B}_\mu v(p_4) T^a_{ij} T^b_{kl}, \\
			i\mathcal{M}_{gg \rightarrow t \bar{t}}^{\text{NC}} 
			&= \epsilon_\mu^a(k_1) \Bigg[
			g_s f^{abc} \left(g_{\mu\nu}(k_1-k_2)_\rho + g_{\nu\rho}(k_2+\ell_2)_\mu - g_{\rho\mu}(\ell_2+k_1)_\nu\right)
			+\\ &\quad+ \frac{1}{2} g_s d^{abc} \Theta_3^{\mu\nu\rho} 
			\Bigg] \epsilon_\nu^b(k_2) 
			 \frac{-i\eta^{\rho\sigma}}{\ell_2^2} \bar{u}(p_3) (-ig_s \mathcal{B}_\sigma T^c) v(p_4) \\
			&\quad + \bar{u}(p_3)\left(ig_s\mathcal{B}^\mu T^a_{ij}\right)\epsilon^a_\mu(k_1)\frac{i(\gamma \cdot \ell_3 + m_{\mathrm{top}})}{\ell_3^2 - m_{\mathrm{top}}^2}\epsilon^b_\nu(k_2) \left(-ig_s \mathcal{B}^\nu T^b_{kl} v(p_4)\right) \\
			&\quad + \bar{u}(p_4)\left(ig_s\mathcal{B}^\mu T^a_{ij}\right)\epsilon^a_\mu(k_1)\frac{i(\gamma \cdot \ell_4 + m_{\mathrm{top}})}{\ell_4^2 - m_{\mathrm{top}}^2}\epsilon^b_\nu(k_2) \left(-ig_s \mathcal{B}^\nu T^b_{kl} v(p_3)\right),
\end{align*} 
where $m_{\mathrm{top}}$ is the top quark mass, and $\mathcal{B}^\mu$ and $\Theta_3^{\mu\nu\rho}$ are coefficients for the quark-antiquark-gluon and three-gluon vertices in the NC-SM \cite{melic2005standard}:
\begin{align*}
			\mathcal{B}^\mu &= \gamma^\mu - \frac{i}{2} \left[ (p_{\mathrm{in}} \Theta p_{\mathrm{out}}) \gamma^\mu - (p_{\mathrm{out}} \Theta)_\mu (\gamma \cdot p_{\mathrm{in}} - m_q) - (\gamma \cdot p_{\mathrm{out}} - m_q)(\Theta p_{\mathrm{in}})_\mu \right], \\
			\Theta_3^{\mu\nu\rho} &= -(k_1 \Theta k_2) \left[(k_1 - k_2)^\rho g^{\mu\nu} + (k_2 - \ell_2)^\mu g^{\nu\rho} + (\ell_2 - k_1)^\nu g^{\rho\mu}\right] \\
			&\quad - \Theta^{\mu\nu}\left[k_1^\rho(k_2 \cdot \ell_2) - k_2^\rho(k_1 \cdot \ell_2)\right] \\
			&\quad - \Theta^{\nu\rho}\left[k_2^\mu(\ell_2 \cdot k_1) - \ell_2^\mu(k_2 \cdot k_1)\right] \\
			&\quad - \Theta^{\rho\mu}\left[\ell_2^\nu(k_1 \cdot k_2) - k_1^\nu(\ell_2 \cdot k_2)\right] \\
			&\quad + (\Theta k_2)^\mu \left[g^{\nu\rho} \ell_2^2 - \ell_2^\nu \ell_2^\rho\right] + (\Theta \ell_2)^\mu \left[g^{\nu\rho} k_2^2 - \ell_2^\nu k_2^\rho\right] \\
			&\quad + (\Theta \ell_2)^\nu \left[g^{\mu\rho} k_1^2 - k_1^\mu k_1^\rho\right] + (\Theta k_1)^\nu \left[g^{\mu\rho} \ell_2^2 - \ell_2^\mu \ell_2^\rho\right] \\
			&\quad + (\Theta k_1)^\rho \left[g^{\mu\nu} k_2^2 - k_2^\mu k_2^\nu\right] + (\Theta k_2)^\rho \left[g^{\mu\nu} k_1^2 - k_1^\mu k_1^\nu\right],
\end{align*} 
where $(k_i \Theta k_j) = k_{i\mu} \Theta^{\mu\nu} k_{j\nu}$, $(\Theta k)^\mu = \Theta^{\mu\nu} k_\nu$, and $(\Theta k)^2 = (\Theta k)^\nu (\Theta k)_\nu$. In this work, we assume space-space non-commutativity, i.e., $\Theta^{i0} = 0$, motivated by unitarity considerations \cite{gomis2000space}, and without loss of generality, we set $\Theta^{21} = \Theta$ and $\Theta^{13} = \Theta^{23} = 0$. Under these assumptions, the color and polarization-averaged/summed squared amplitude in our kinematics is found to be:
\begin{align*}
			|\mathcal{M}^{\text{NC}}_{q\bar{q} \rightarrow t \bar{t}}|^2 &= |\mathcal{M}_{q\bar{q} \rightarrow t \bar{t}}|^2, \\
			|\mathcal{M}^{\text{NC}}_{gg \rightarrow t \bar{t}}|^2 &= |\mathcal{M}_{gg \rightarrow t \bar{t}}|^2 - \frac{N g_s^4 C_F}{64} \left(1 - \frac{4}{N^2}\right) \Theta^2 
			\left[ (u-t)^2 - s(s + 2m_{\mathrm{top}}^2) \right],
\end{align*}
where $C_F = 4/3$ is the Casimir scalar in the fundamental representation of SU(3), $N=3$ is the dimension of SU(3), and $s$, $t$, and $u$ are the Mandelstam variables. The quantities $|\mathcal{M}_{q\bar{q} \rightarrow t \bar{t}}|^2$ and $|\mathcal{M}_{gg \rightarrow t \bar{t}}|^2$ are the color and polarization-averaged/summed squared amplitudes for the $q\bar{q}$ and $gg$ channels, respectively, in the SM \cite{ParticleDataGroup:2022pth}.

\section{Single Top Quark Production in NC-SM}
\label{sec:single_top_NCSM}
	
\begin{figure}[ht]
		\centering
		\includegraphics[width=0.25\textwidth]{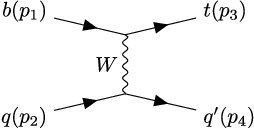}
		\includegraphics[width=0.25\textwidth]{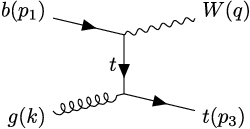}
		\includegraphics[width=0.25\textwidth]{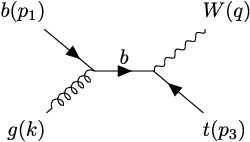}
		\caption{Leading-order diagrams for $gb \rightarrow tW^{-}$ \cite{kidonakis2010two} and $qb \rightarrow q't$ \cite{sirunyan2017cross}.}
		\label{fig02}
	\end{figure}

In the NC-SM, single top quark production via the $t$-channel and $tW$-associated channel proceeds through the same Feynman diagrams as in the SM. The leading-order diagrams for these processes are shown in Fig.~\ref{fig02}.
	
\begin{figure}[ht]
		\centering
		\includegraphics[width=0.23\textwidth]{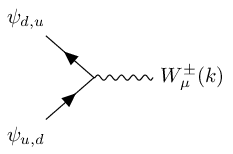}\quad
		\begin{minipage}[c]{0.65\textwidth}
			\centering
			$\displaystyle
			\frac{ie}{2\sqrt{2}\sin\theta_W}
			\begin{pmatrix}
				V_{ud} \\
				V^*_{ud}
			\end{pmatrix}
			\Big\{
			\Big[\gamma_\mu - \frac{i}{2}\Theta_{\mu\nu\rho}k^\nu p_{\text{in}}^\rho\Big](1-\gamma_5) -
			\frac{1}{2}\Theta_{\mu\nu}\Big[
			\begin{pmatrix}
				m_u \\ m_d
			\end{pmatrix}
			p_{\text{in}}^\nu(1-\gamma_5) - 
			\begin{pmatrix}
				m_d \\ m_u
			\end{pmatrix}
			p^\nu_{\text{out}}(1+\gamma_5)
			\Big]
			\Big\}
			$
		\end{minipage}
		\caption{Electroweak vertices modified by non-commutative geometry (NCG) \cite{melic2005standard}.}
		\label{fig03}
	\end{figure}

In this section, in addition to the gluon-quark-quark vertex, we consider the $W$ boson-quark-quark vertex, which is modified by NCG at leading order in the non-commutative parameter $\Theta$. This modification introduces a correction term, as shown in Fig.~\ref{fig03}.
	
In Fig.~\ref{fig03}, $m_u$ and $m_d$ represent the masses of the up-type and down-type quarks, respectively, $V_{ud}$ denotes the CKM matrix element, and $\theta_W$ is the Weinberg angle. The color- and polarization-averaged/summed squared amplitudes for the two channels, using the same kinematics and non-commutative parameter $\Theta$ as in the previous section, are given by:
\begin{align*}
		|\mathcal{M}_{\text{$t$-channel}}^{\text{NC}}|^2 &= |\mathcal{M}_{\text{$t$-channel}}^{\text{SM}}|^2 - \frac{g_W^2 |V_{qq'}|^2 |V_{tb}|^2}{64 (t-m_W^2)} \Theta^2 p_t^2 m_{\mathrm{top}}^2 t(t-m_{\mathrm{top}}^2), \\
		|\mathcal{M}_{\text{$tW$-channel}}^{\text{NC}}|^2 &= |\mathcal{M}_{\text{$tW$-channel}}^{\text{SM}}|^2 + \frac{g_s^2 g_W^2 C_F C_A |V_{tb}|^2}{16 N (N^2-1) m_W^2} \left[
		p_t^2 \Theta \chi_1 +
		p_t^2 \Theta^2 \left(\chi_2^s + \chi_2^t + \chi_2^{st}\right)
		\right],
\end{align*}
where
\begin{align*}
		\chi_1 &= \frac{4(m_{\mathrm{top}}^2 + m_W^2 - s)s^2}{s^2} + \frac{8(t-2m_W^2)(t-m_{\mathrm{top}}^2)}{(t-m_{\mathrm{top}}^2)^2} + \frac{8s(s+m_W^2)(t-m_{\mathrm{top}}^2)}{(t-m_{\mathrm{top}}^2)s}, \\
		\chi_2^s &= \frac{1}{s^2} \left[ m_W^4 \left(u m_{\mathrm{top}}^2 - 4 m_{\mathrm{top}}^4 + s^2\right) + t \left(m_{\mathrm{top}}^4 + s^2\right) \left(s-m_{\mathrm{top}}^2\right) \right. \\
		& \quad \left. + m_W^2 m_{\mathrm{top}}^4 (-s + 3t + u) + 2 s t m_{\mathrm{top}}^2 m_W^2 - s^2 m_W^2 (s + t - u) \right], \\
		\chi_2^t &= \frac{1}{(t-m_{\mathrm{top}}^2)^2} \left[
		2 m_{\mathrm{top}}^2 m_W^2 (st + 2 m_{\mathrm{top}}^4 - m_{\mathrm{top}}^2(3s + 2u)) \right. \\
		& \quad \left. + (t-m_{\mathrm{top}}^2) \left(m_{\mathrm{top}}^2 \left(2 s m_W^2 - 2 m_W^4 + t u\right) + t \left(2 u m_W^2 + s t\right) - m_{\mathrm{top}}^4 m_W^2\right)
		\right], \\
		\chi_2^{st} &= \frac{1}{(t-m_{\mathrm{top}}^2) s} \left[
		t\left(-m_{\mathrm{top}}^4 + s^2 + t^2 - u^2\right) \right. \\
		& \quad \left. + m_W^2 \left(m_W^2(6s + t) + 2(s - t)m_{\mathrm{top}}^2 - 2(m_W^4 + m_{\mathrm{top}}^4) + u(u - s)\right)
		\right],
\end{align*}
where $|\mathcal{M}_{\text{$t$-channel}}^{\text{SM}}|^2$ and $|\mathcal{M}_{\text{$tW$-channel}}^{\text{SM}}|^2$ are the color- and polarization-averaged/summed squared amplitudes in the SM \cite{harris2002fully}:
\begin{align*}
		|\mathcal{M}_{\text{$t$-channel}}^{\text{SM}}|^2 &= \frac{|V_{q\bar{q}'}|^2 |V_{t\bar{b}}|^2 g_W^2}{4(t-m_W^2)^2} s(s-m_{\mathrm{top}}^2), \\
		|\mathcal{M}_{\text{$tW$-channel}}^{\text{SM}}|^2 &= \frac{|V_{t\bar{b}}|^2 g_W^2 g_s^2 C_F}{N(N^2-1)} \Bigg[\frac{2s(u-m_{\mathrm{top}}^2)}{s^2} + \\&+\frac{m_{\mathrm{top}}^2(s-2t+u) - m_{\mathrm{top}}^4 + 2tu}{(t-m_{\mathrm{top}}^2)^2} - \frac{2s(2m_{\mathrm{top}}^2+t)}{s(t-m_{\mathrm{top}}^2)}\Bigg].
\end{align*}

\section{Numerical Results and Analysis}

In this section, we present a detailed numerical analysis of top quark pair production and single top quark production within the framework of the NCSM. We focus on how the NC parameter $\Theta$ can be probed using fully inclusive cross-sections and differential distributions of key observables. The total cross-section in the NCSM framework, incorporating NC corrections, is expressed as
\begin{equation}
	\sigma^{\mathrm{NC}} = \sum_{\delta} \int dx_a \, dx_b \, f_a(x_a, \mu_\mathrm{F}^2) \, f_b(x_b, \mu_\mathrm{F}^2) \, \Xi \, \frac{dp_t^2 \, dy}{16\pi^2 s} \, \delta(s + t + u - m_{\rm{top}}^2-m_{X}^2) \, \overline{\left|\mathcal{M}^{\mathrm{NC}}_{\delta}\right|^2}\,,
	\label{eq:sigmaNC}
\end{equation}
where $x_a$ and $x_b$ are the momentum fractions of the incoming partons, $f_a(x_a, \mu_\mathrm{F}^2)$ and $f_b(x_b, \mu_\mathrm{F}^2)$ are the parton distribution functions (PDFs), $\Xi$ is a process-dependent kinematic factor, and $\overline{\left|\mathcal{M}^{\mathrm{NC}}_{\delta}\right|^2}$ is the squared amplitude averaged over initial states and summed over final states for the specific channel $\delta$.

The numerical analysis is performed by generating parton-level events for the process $pp \to t + X$ ($X = \bar{t}, W^-, j$) at leading order (LO) within the SM using the Monte Carlo event generator \texttt{MadGraph5\_aMC@NLO}~\cite{Maltoni:2002qb, Alwall:2014hca}. Non-commutative geometry  corrections are incorporated into the event weights according to the following formula
\begin{align}
	\label{eq:wt}
	\sigma^{\texttt{NC}}=\frac{\sigma^{\texttt{SM}}}{N}\sum_{\text{events}}(\overline{|\mathcal{M}_\delta^{\texttt{NC}}|^2}/\overline{|\mathcal{M}_\delta^{\texttt{SM}}|^2})
\end{align}
where the sum is over all $N$ events in the generated sample, $\sigma^{\texttt{SM}}$ is the SM cross-section obtained from \texttt{MadGraph5\_aMC@NLO}, and $\delta$ represents the specific production channel. The kinematics for each event are accessed using \texttt{MadAnalysis5}~\cite{Conte:2012fm}, and the corresponding weights are calculated as shown in Eq.~\eqref{eq:wt}.

For all quantitative results, we consider $pp$ collisions at $\sqrt{s} = 14$ TeV and set the top quark pole mass to $m_t = 173.0$ GeV. The parton distribution functions (PDFs) are taken from NNPDF3.1 NNLO~\cite{ball2017parton} via \texttt{LHAPDF}~\cite{buckley2015lhapdf6}, and we impose a rapidity cut of $\eta < 5$ for jets. The renormalization and factorization scales are set to $\mu_\mathrm{F} = \mu_\mathrm{R} = m_\mathrm{top}$ for our analysis.

To provide a more accurate assessment of the effects of NC geometry, we compare NC distributions with fixed-order distributions (LO, NLO, and NNLO) within the SM. The fixed-order Monte Carlo program \texttt{MCFM}~\cite{Campbell:2019dru} is used to compute transverse momentum and rapidity distributions at NLO. Scale uncertainties are evaluated by varying $\mu_\mathrm{R} = \mu_\mathrm{F} = m_\mathrm{top}$ simultaneously by factors of $2$ and $1/2$ to assess the sensitivity of the results to these unphysical scales.

\subsection{Inclusive Cross-section}
	
\begin{figure}[ht]
		\centering
		\includegraphics[width=0.32\textwidth]{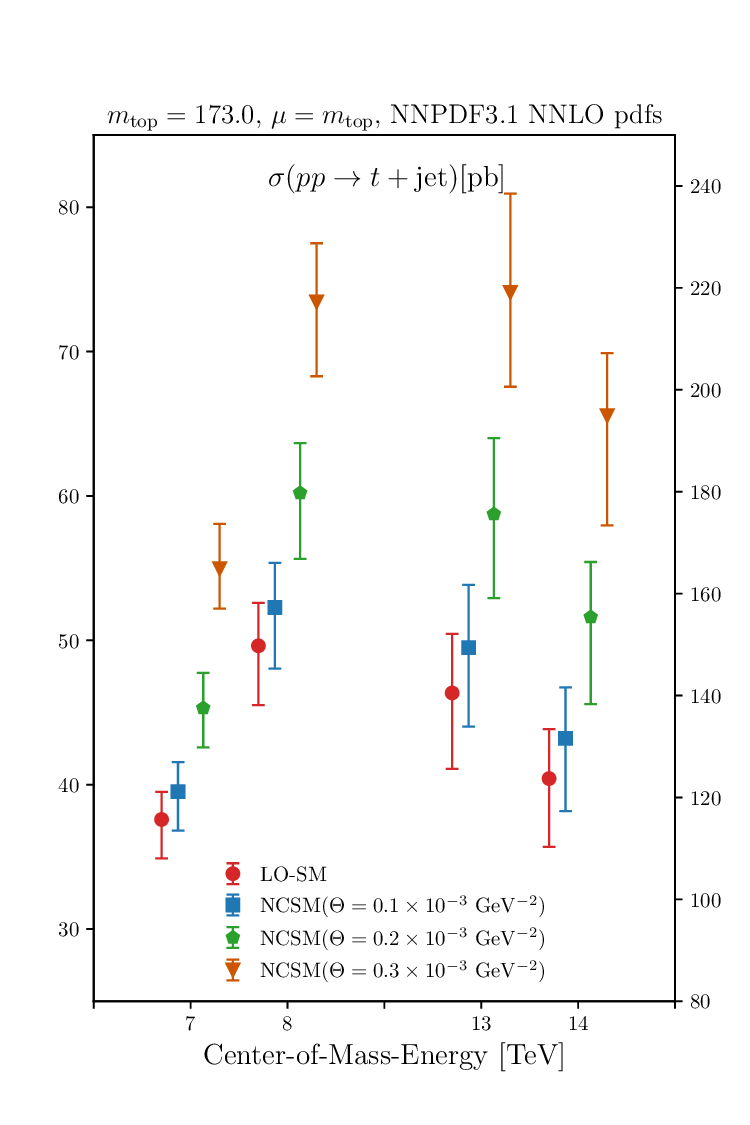}
		\includegraphics[width=0.32\textwidth]{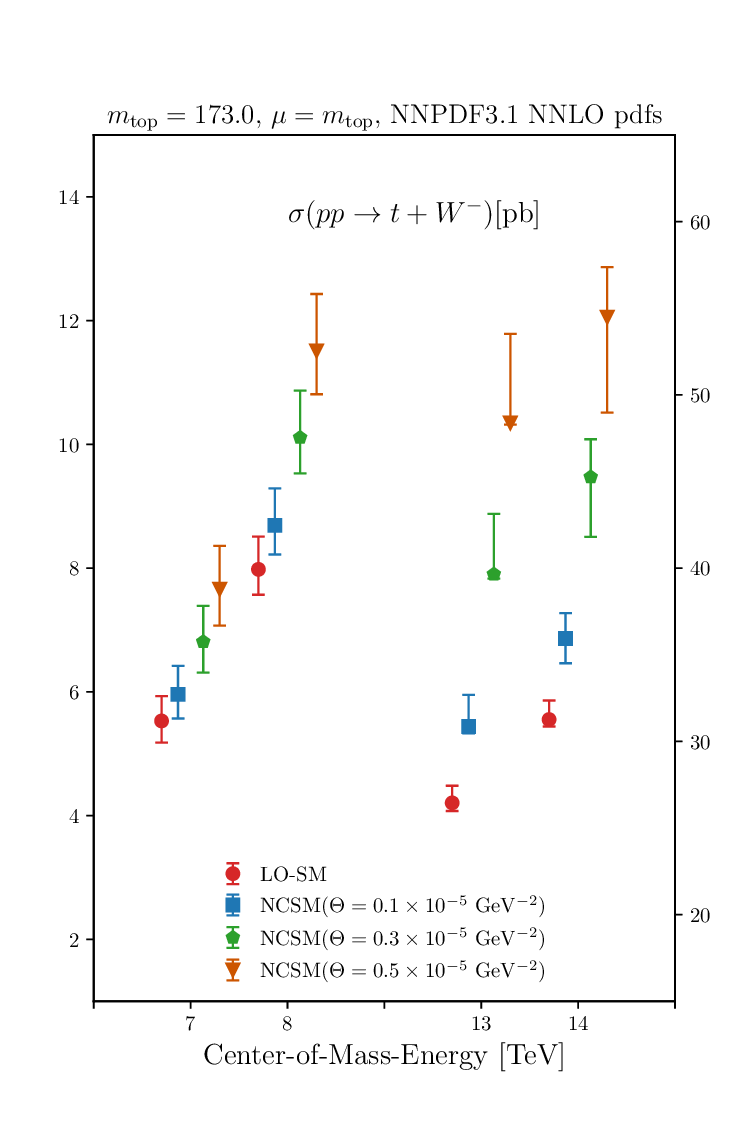}
		\includegraphics[width=0.32\textwidth]{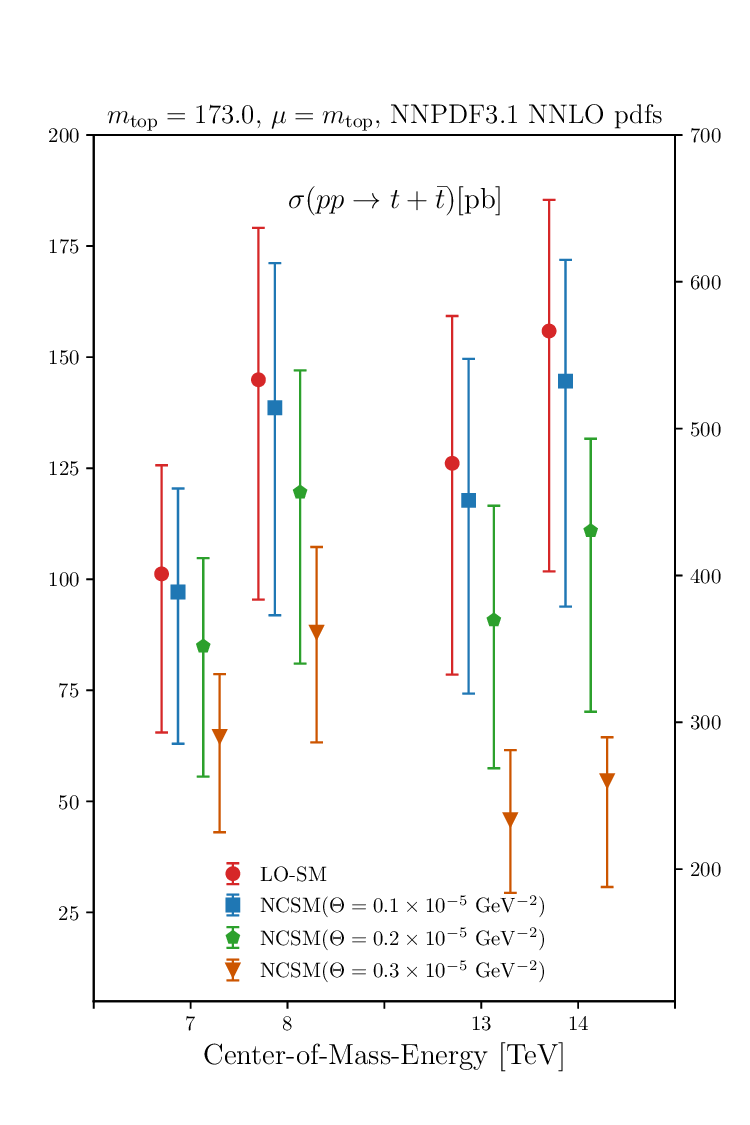}
		\vspace*{8pt}
		\caption{Inclusive cross sections for $pp \to t + X$ ($X = \bar{t}, W^-, j$) at LO within SM and NCSM for different values of the NC parameter $\Theta$ at the LHC. The error bars represent scale uncertainties obtained by varying the renormalization and factorization scales from $\mu_\mathrm{F} = \mu_\mathrm{R} = m_\mathrm{top}/2$ to $2m_\mathrm{top}$. \protect\label{fig12}}
	\end{figure}	
In Fig.~\ref{fig12}, we plot the inclusive cross-sections for top quark production with $X = \bar{t}, W^-, \mathrm{jet}$ at $\sqrt{s} = 7$, $8$, $13$, and $14$ TeV. The left vertical axis corresponds to the lower center-of-mass energies, while the right vertical axis corresponds to the higher energies. Scale variations are calculated by setting the scales in top quark and $X$ production to be the same and varying them simultaneously. The error bars represent perturbative scale variations at different orders. 
	
The incorporation of NC corrections leads to an increase in scale variations, which becomes particularly pronounced for specific values of $\Theta$, such as $0.3 \times 10^{-5}$ GeV$^{-2}$, $0.3 \times 10^{-3}$ GeV$^{-2}$, and $0.5 \times 10^{-5}$ GeV$^{-2}$ for $X = \bar{t}$, jet, and $W^{-}$, respectively. As expected, the intrinsic theoretical complexity in non-commutative geometry leads to greater prediction uncertainties, resulting in increased sensitivity of the theoretical calculations to scale variations.
	
\begin{table}
	\centering
	\caption{Cross-section for the production of top quark with $X (X=\bar{t}, W^-, j)$ in NC-geometry for different choices of the $\Theta$ parameter.  At the LHC with
		different center of mass energies. \label{tab01}}
	\begin{tabular}{c|c|c|c|c}
		\hline
		\multicolumn{5}{|c|}{pair top quark production}                                                                                                      \\ \hline
		NC parameter $\Theta$ & \multicolumn{4}{c}{ NC cross-section $\sigma^{\texttt{NC}}$[pb]}                                                         \\ \cline{2-5} 
		$\times 10^{-5}$ GeV$^{-2}$)          & 7 TeV & 8 TeV& 13 TeV & 14 TeV \\ \hline
		0.0 ($\sigma^{\texttt{SM}}$)                         &$101.2^{-24.1\%}_{+35.2\%}$ & $144.9^{-23.6\%}_{+34.1\%}$& $476.4^{-21.1\%}_{+30.2\%}$  & $566.5^{-15.7\%}_{+28.8\%}$  \\ 
		stat. unc.& $\pm 0.03$&$\pm0.05$&$\pm0.16$&$\pm0.24$\\ \hline
		$0.1$                           & $138.6^{-23.4\%}_{+33.6\%}$ & $97.13^{-23.9\%}_{+35.1\%}$& $451.2^{-21.3\%}_{+29.2\%}$  & $532.4^{-15.5\%}_{+28.9\%}$  \\
		0.2                           & $119.6^{-22.9\%}_{+32.3\%}$ & $84.93^{-23.3\%}_{+34.6\%}$ & $369.6^{-21.1\%}_{+27.3\%}$  & $430.4^{-14.6\%}_{+28.6\%}$  \\ 
		0.3                           & $88.08^{-21.8\%}_{+28.2\%}$ & $64.6^{-21.7\%}_{+33.4\%}$  & $233.6^{-20.3\%}_{+21.3\%}$  & $260.3^{-11.3\%}_{+27.9\%}$  \\ \hline
		\multicolumn{5}{|c|}{single top quark production-$t$ channel}                                                                                                      \\ \hline
		NC parameter $\Theta$ & \multicolumn{4}{c}{ NC cross-section $\sigma^{\texttt{NC}}$[pb]}                                                         \\ \cline{2-5} 
		$\times 10^{-3}$ GeV$^{-2}$)          & 7 TeV & 8 TeV& 13 TeV & 14 TeV \\ \hline
		0.0 ($\sigma^{\texttt{SM}}$)                          &$37.59^{+5.1\%}_{-7.2\%}$ & $49.62^{+5.9\%}_{-8.3\%}$& $123.7^{+7.9\%}_{-10.8\%}$  & $140.5^{+8.3\%}_{-10.6\%}$  \\ 
		stat. unc.& $\pm 0.07$&$\pm0.1$&$\pm0.3$&$\pm0.013$\\\hline
		0.1                           & $39.52^{+5.2\%}_{-6.8\%}$ & $52.27^{+5.9\%}_{-8.1\%}$& $131.6^{+7.6\%}_{-10.9\%}$  & $149.4^{+8.2\%}_{-10.4\%}$  \\
		0.2                           & $45.31^{+5.4\%}_{-6.1\%}$ & $60.21^{+5.7\%}_{-7.6\%}$ & $155.4^{+6.9\%}_{-11.0\%}$  & $175.6^{+8.5\%}_{-9.4\%}$  \\ 
		0.3                           & $54.96^{+5.6\%}_{-5.0\%}$ & $73.44^{+5.5\%}_{-7.0\%}$ & $194.9^{+6.3\%}_{-11.0\%}$  & $219.1^{+8.9\%}_{-8.4\%}$ \\ \hline
		\multicolumn{5}{|c|}{single top quark production- W associated channel}                                                                                                      \\ \hline
		NC parameter $\Theta$ & \multicolumn{4}{c}{ NC cross-section $\sigma^{\texttt{NC}}$[pb]}                                                         \\ \cline{2-5} 
		$\times 10^{-5}$ GeV$^{-2}$)          & 7 TeV & 8 TeV& 13 TeV & 14 TeV \\ \hline
		0.0 ($\sigma^{\texttt{SM}}$)                      &$5.53^{-7.2\%}_{+6.3\%}$& $7.98^{-6.6\%}_{+5.1\%}$& $26.45^{-3.7\%}_{+1.8\%}$  & $31.26^{-3.5\%}_{+1.2\%}$  \\ 
		stat. unc.& $\pm 0.002$&$\pm0.003$&$\pm0.009$&$\pm0.01$\\ \hline
		0.1                           & $5.96^{-7.7\%}_{+6.5\%}$ & $8.69^{-6.9\%}_{+5.4\%}$& $30.86^{-5.8\%}_{+1.3\%}$  & $35.94^{-4.1\%}_{+3.9\%}$  \\
		0.3                           & $6.81^{-8.5\%}_{+7.3\%}$ & $10.11^{-7.5\%}_{+5.7\%}$ & $39.65^{-8.8\%}_{+0.7\%}$  & $45.26^{-4.8\%}_{+7.6\%}$  \\ 
		0.5                           & $7.66^{-9.1\%}_{+7.7\%}$ & $11.51^{-7.9\%}_{+6.8\%}$ & $48.38^{-10.6\%}_{+0.2\%}$  & $54.49^{-5.3\%}_{+10.1\%}$  \\ \hline
	\end{tabular}
\end{table}
	
Table \ref{tab01} shows the numerical values of the cross section in pb for a top quark mass of $173.0$ GeV in the NCSM for different choices of $\Theta$ and various collision energies at the LHC. We observe that the non-commutative corrections can either increase or decrease the cross-section value. For example, in the process of producing a top quark pair at a center-of-mass energy of 14 TeV, the cross-section value decreases by 54\% for $\Theta = 0.3 \times 10^{-5}$ GeV$^{-2}$. In contrast, for the same value of the NC parameter, the cross-section for single top quark production increases by 55\% through the $t$-channel and by 57\% through the $tW$ associated production channel.

\subsection{Differential Distribution Analysis}

In this section, we continue to evaluate the effect of non-commutative geometry on the process of top quark production with $X$ ($X=\bar{t}, W^-, \mathrm{jet}$). This time, we do so by plotting the transverse momentum distributions and rapidity distributions.

\subsubsection{Top Quark Pair Production}
We first consider differential distributions in $p_t$ and rapidity of the top quark in pair production.
	
\begin{figure}[ht]
		\centering
		\includegraphics[width=0.43\textwidth]{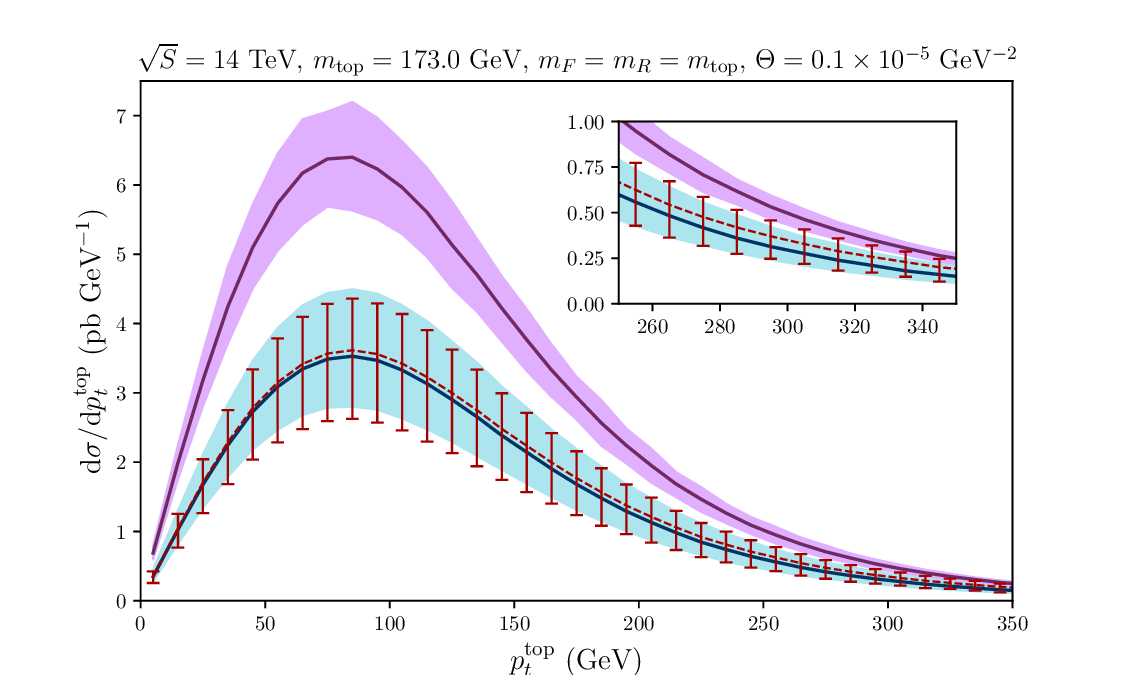}
		\includegraphics[width=0.43\textwidth]{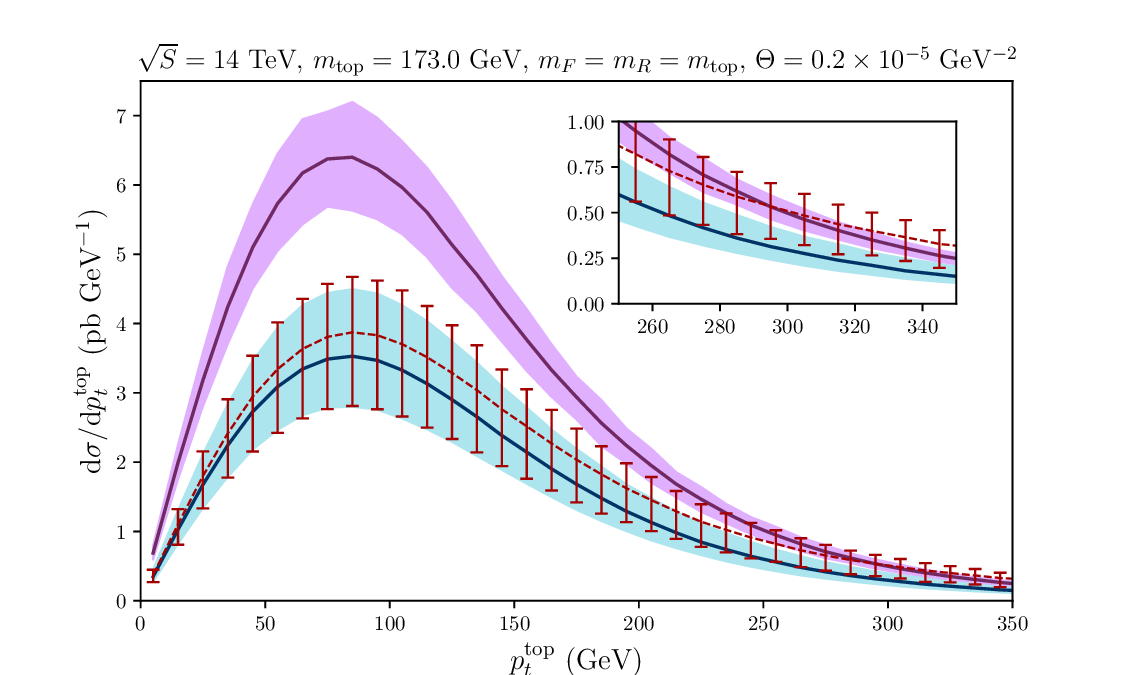}
		\includegraphics[width=0.43\textwidth]{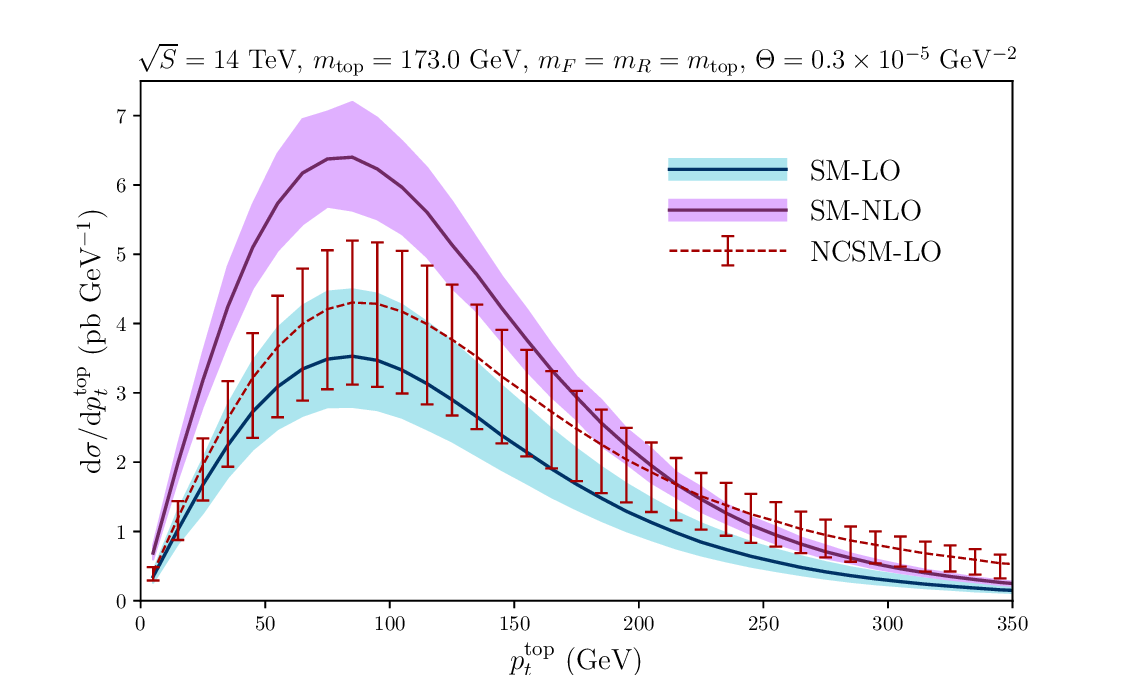}
		\includegraphics[width=0.43\textwidth]{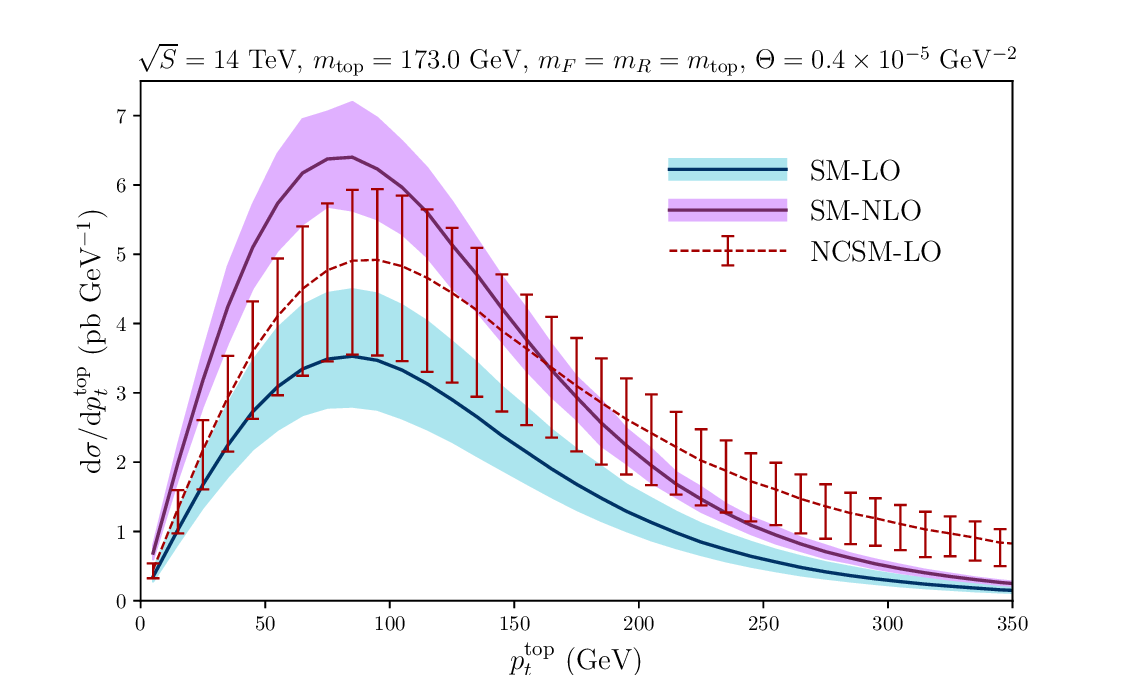}
		\vspace*{8pt}
		\caption{Top quark $p_t^{\rm{top}}$ distribution within the pair production process. \protect\label{fig2}}
\end{figure}

In Fig.~\ref{fig2}, we show the SM-LO, SM-NLO, and NCSM-LO top-quark transverse momentum distributions, $\mathrm{d}\sigma/\mathrm{d}p_t^{\rm{top}}$, at 14~TeV center-of-mass energy. We observe that the effects of non-commutative geometry are significant at high \(p_t^{\rm{top}}\) values. These effects become almost negligible at low \(p_t^{\rm{top}}\) values. Additionally, we notice that for $\Theta > 0.2 \times 10^{-5} \, \mathrm{GeV}^{-2}$, the distributions resulting from non-commutative geometry are clearly separated from the fixed-order distributions at NLO.
	
\begin{figure}[ht]
		\centering
		\includegraphics[width=0.43\textwidth]{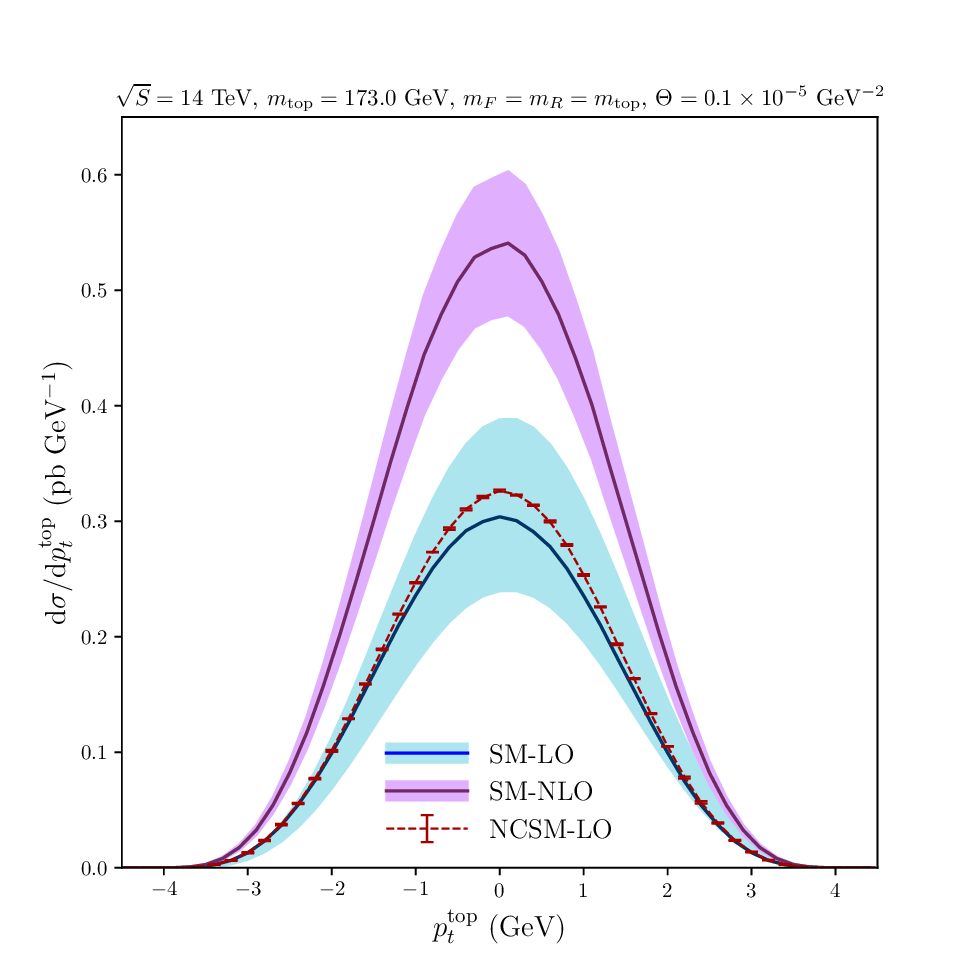}
		\includegraphics[width=0.43\textwidth]{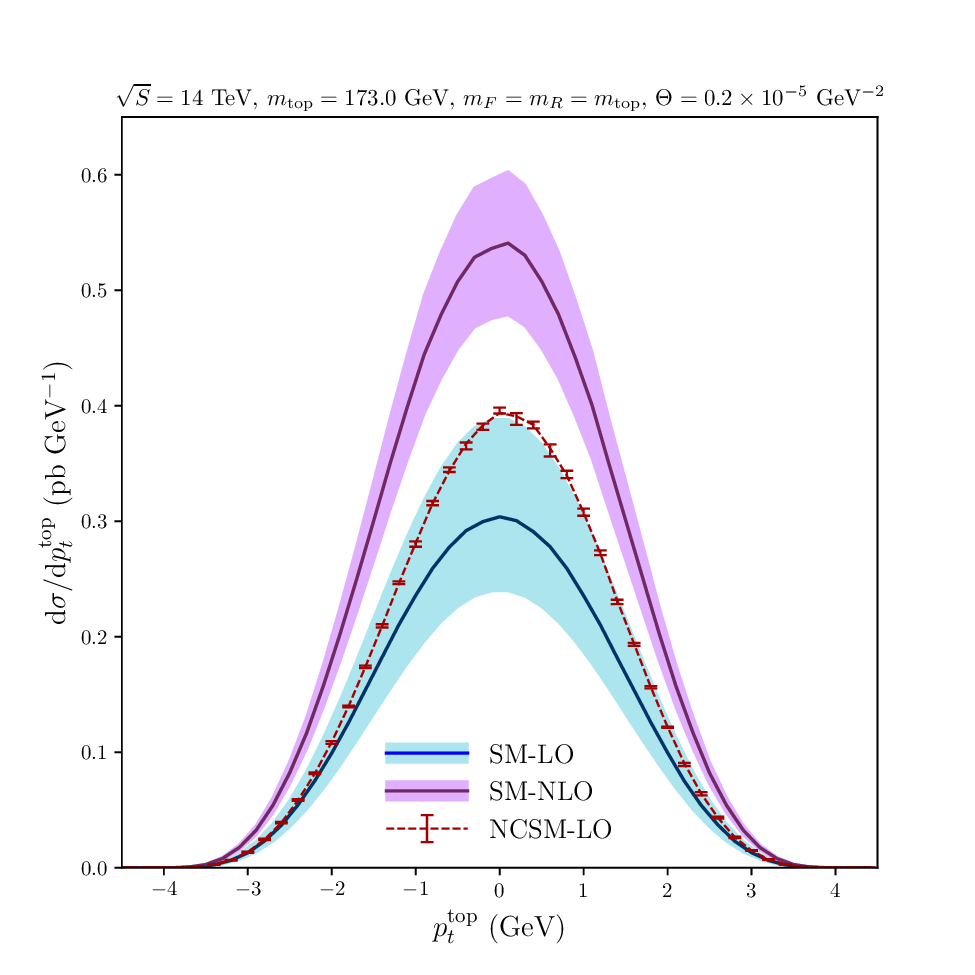}
		\includegraphics[width=0.43\textwidth]{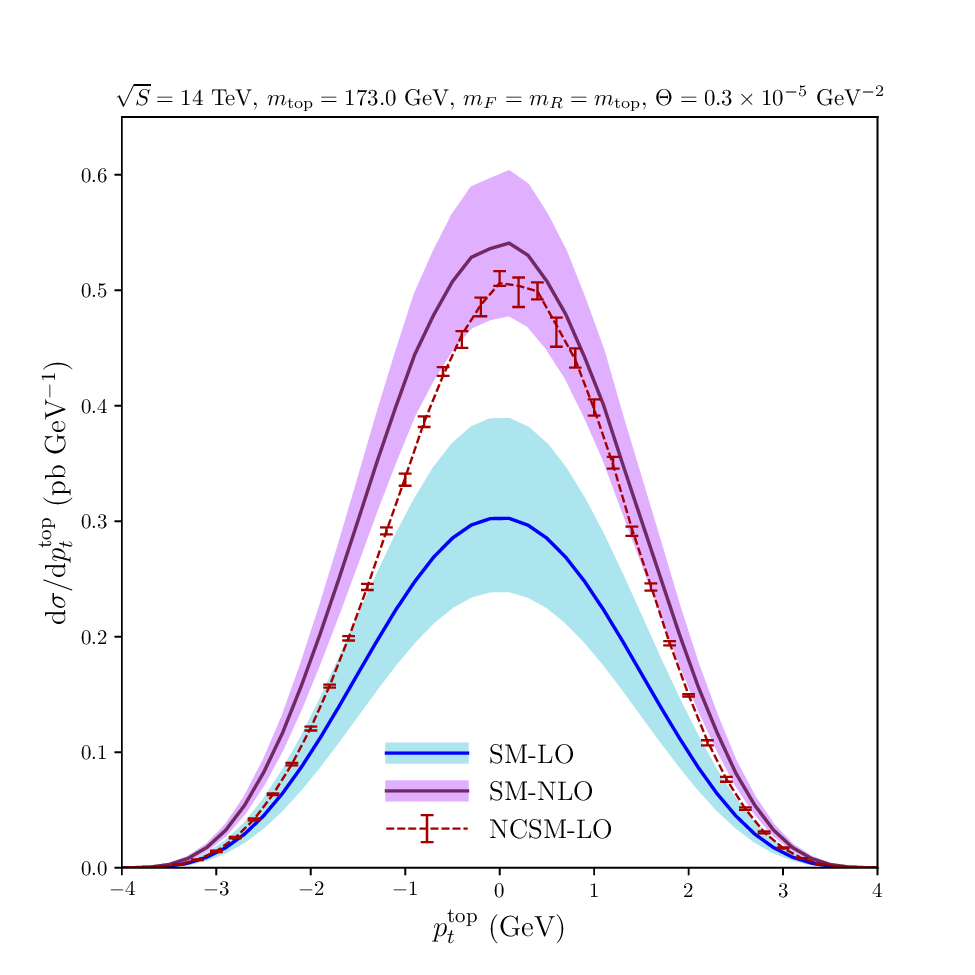}
		\includegraphics[width=0.43\textwidth]{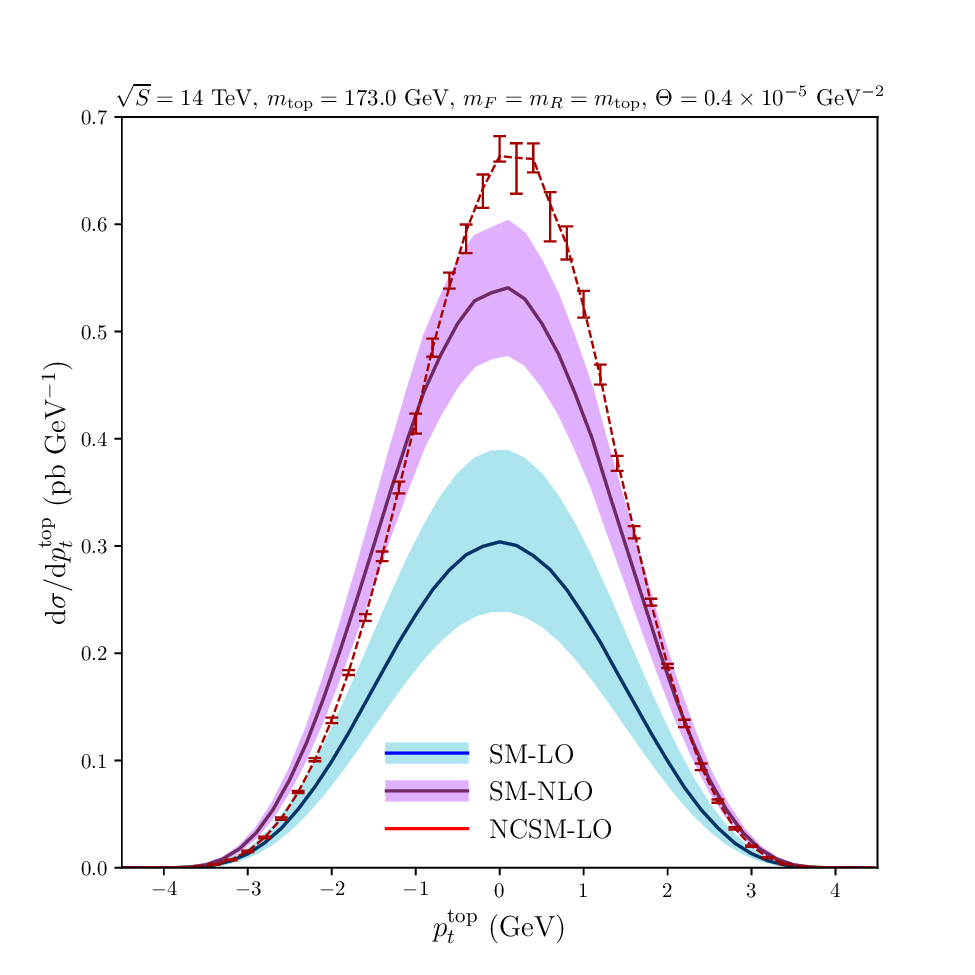}
		\vspace*{8pt}
		\caption{Top quark rapidity distribution. \protect\label{fig3}}
\end{figure}

The top quark rapidity distributions, $1/\sigma \, \mathrm{d}\sigma/\mathrm{d}y$, at 14~TeV center-of-mass energy within the SM at LO and NLO, and within NCSM at LO, are displayed in Fig.~\ref{fig3}. As evident from the figure, NC corrections preserve the shape of the distribution, with significant effects notably in the central region, while being negligible at the distribution edges. The non-commutative distributions exceed the fixed-order distribution at NLO for $\Theta > 0.4 \times 10^{-5} \, \mathrm{GeV}^{-2}$.
	
\subsubsection{Single Top Quark Production}
	
\begin{figure}[ht]
		\centering
		\includegraphics[width=0.43\textwidth]{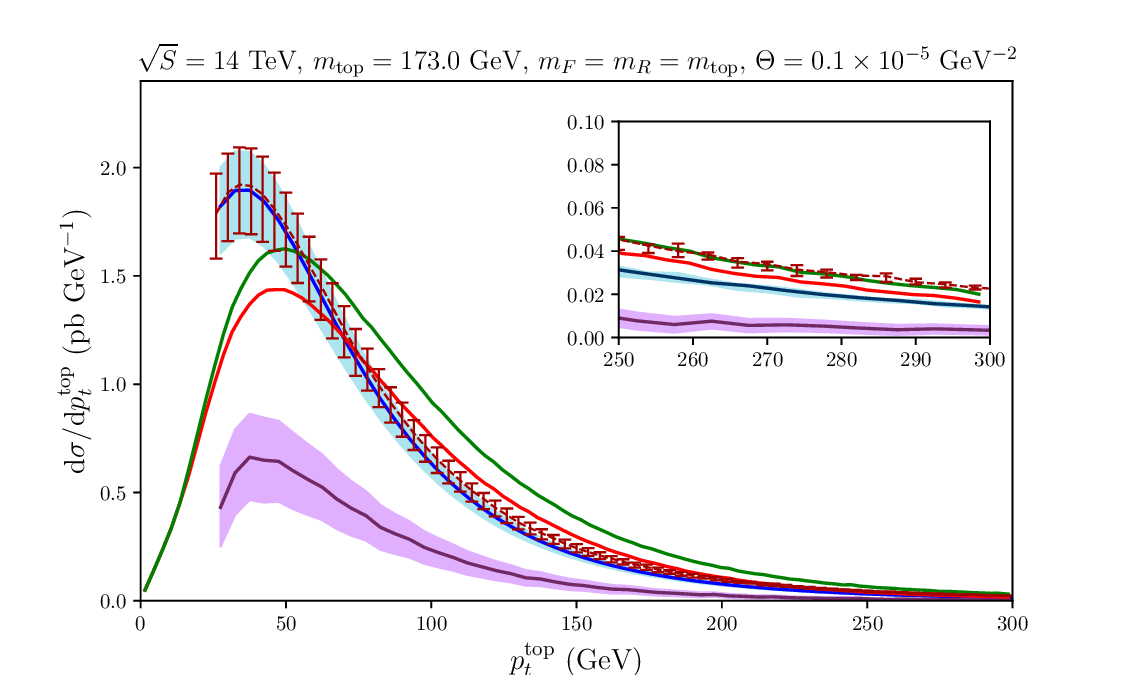}
		\includegraphics[width=0.43\textwidth]{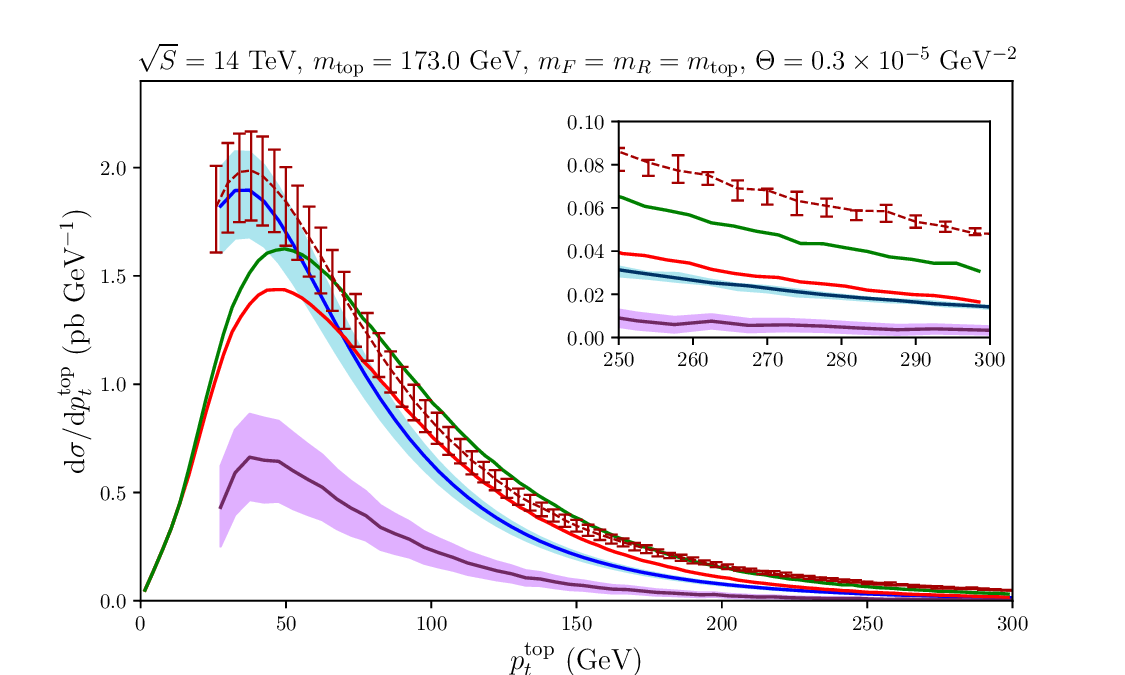}
		\includegraphics[width=0.43\textwidth]{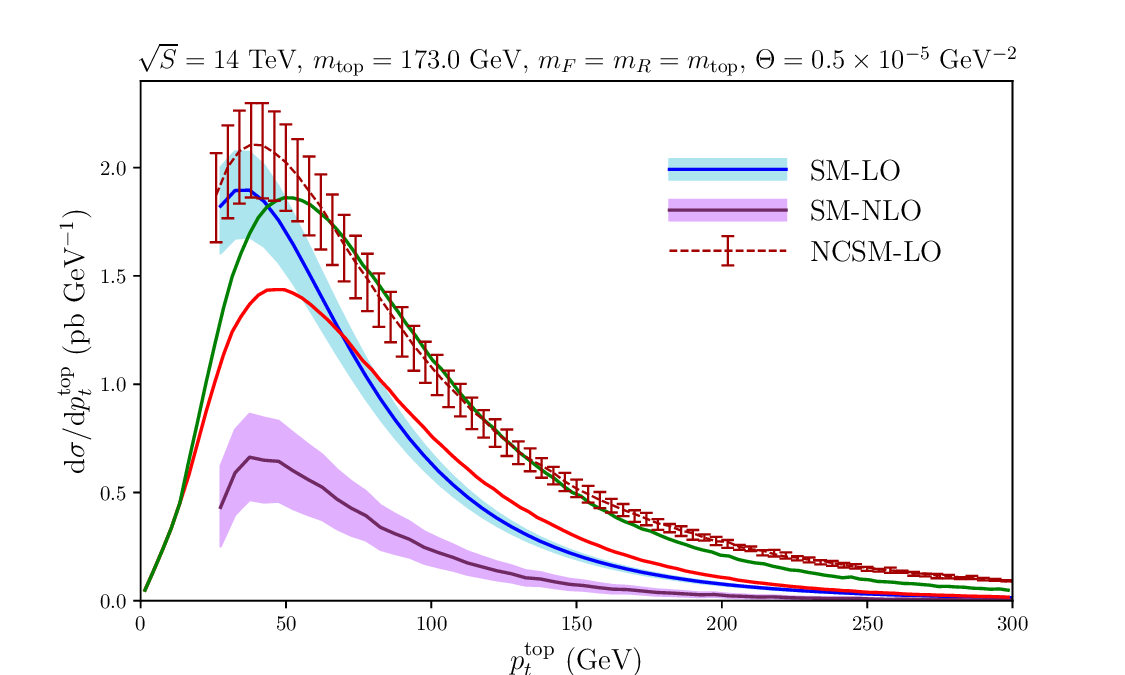}
		\includegraphics[width=0.43\textwidth]{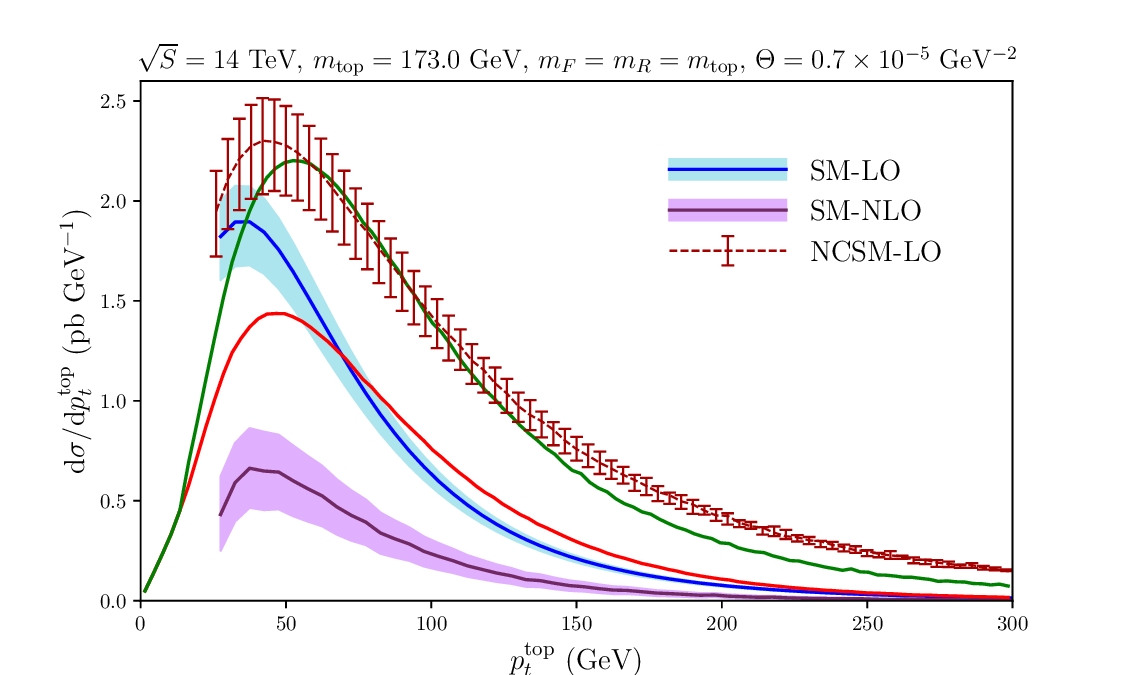}
		\vspace*{8pt}
		\caption{Top quark $p_t^{\rm{top}}$ distribution within single top production in the $t$-channel. \protect\label{fig4}}
\end{figure}
We begin our discussion of single top quark production in the NCSM framework with the $t$-channel transverse momentum distribution shown in Fig.~\ref{fig4}. As shown in the figure, fixed-order distributions (LO and NLO) in both the SM and the NCSM exhibit divergences at small $p_t^{\rm{top}}$. Although the effects of NC geometry do not cover this region of the distribution according to previous results, addressing these divergences is crucial for improving the accuracy of the distributions for comparison with experimental results. To eliminate these soft/collinear divergences, we first observe that they are absent in the NC corrections, indicating that their source lies in the large logarithms of the SM part. Thus, we use two steps to deal with them. First, since the cause of these soft/collinear divergences is the large logarithms, we can perform a resummation at NLL (Next-to-Leading Logarithm) accuracy. We perform resummation on the SM part using Monte Carlo Parton Shower results. Then, we use the following matching formula to eliminate the divergences in the non-commutative distribution
\begin{equation}
		\frac{\mathrm{d}\sigma_{\mathrm{matched}}^\mathrm{NC}}{\mathrm{d} p_t} = \frac{\mathrm{d}\sigma^\mathrm{SM}_{\mathrm{resummed}}}{\mathrm{d} p_t} - \frac{\mathrm{d}\sigma_\mathrm{LO}^\mathrm{SM}}{\mathrm{d} p_t} + \frac{\mathrm{d}\sigma_\mathrm{LO}^\mathrm{NC}}{\mathrm{d} p_t}.
\end{equation}
	
As shown in Fig.~\ref{fig4}, we observe a convergent and synchronized behavior for both the SM parton shower \texttt{Pythia8} and NC-matched top $p_t^{\rm{top}}$ distributions in the low $p_t^{\rm{top}}$ region. However, in the high $p_t^{\rm{top}}$ region, the NC-matched distribution approaches the LO distribution and deviates significantly from the SM predictions.
	
\begin{figure}[ht]
		\centering
		\includegraphics[width=0.43\textwidth]{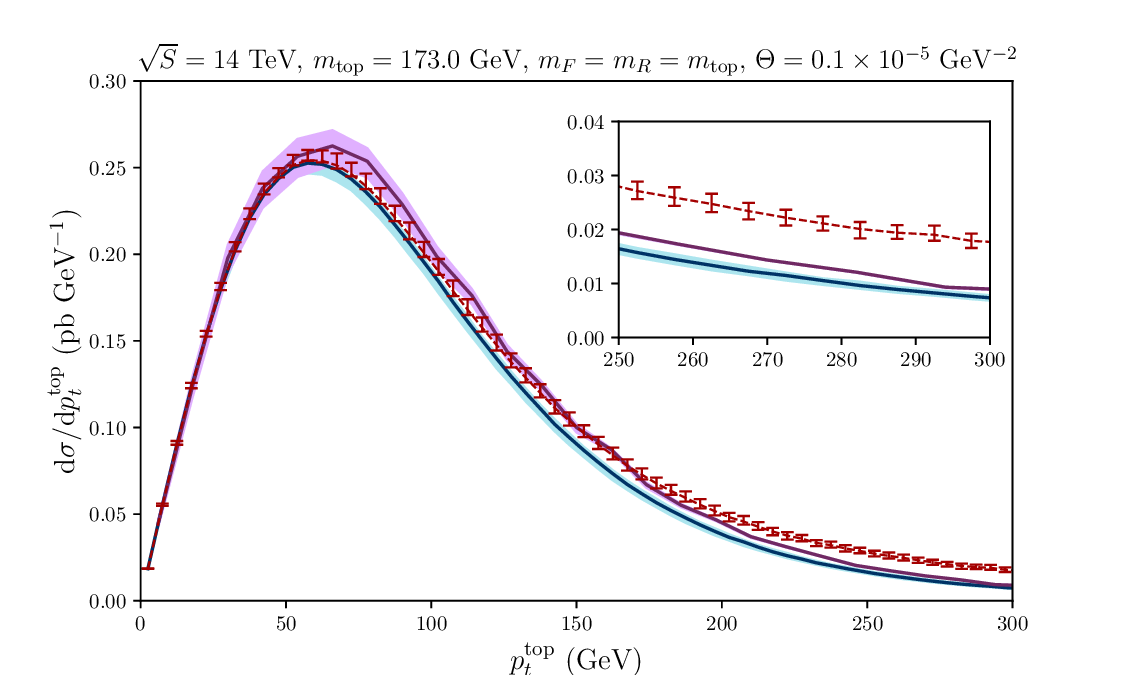}
		\includegraphics[width=0.43\textwidth]{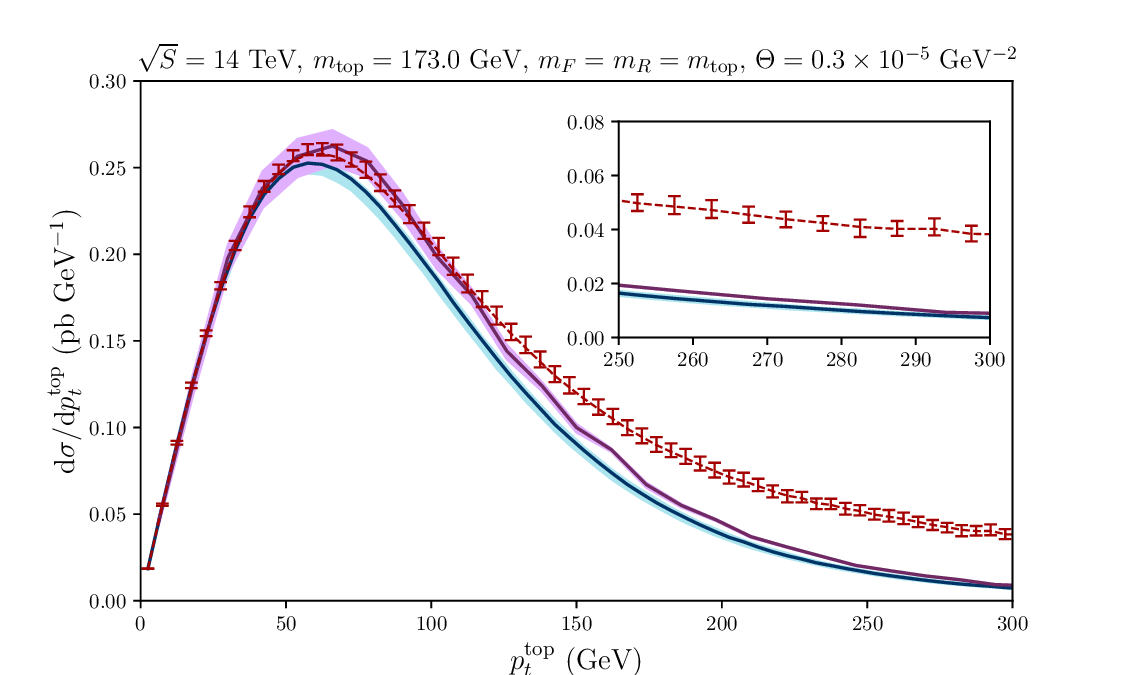}
		\includegraphics[width=0.43\textwidth]{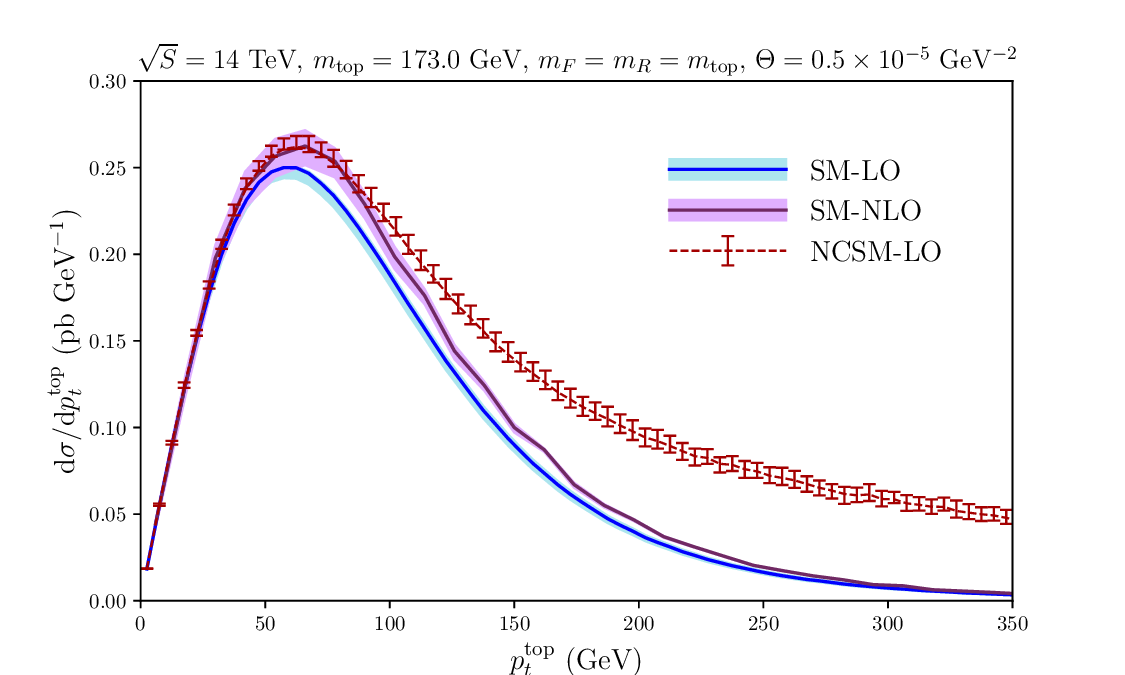}
		\includegraphics[width=0.43\textwidth]{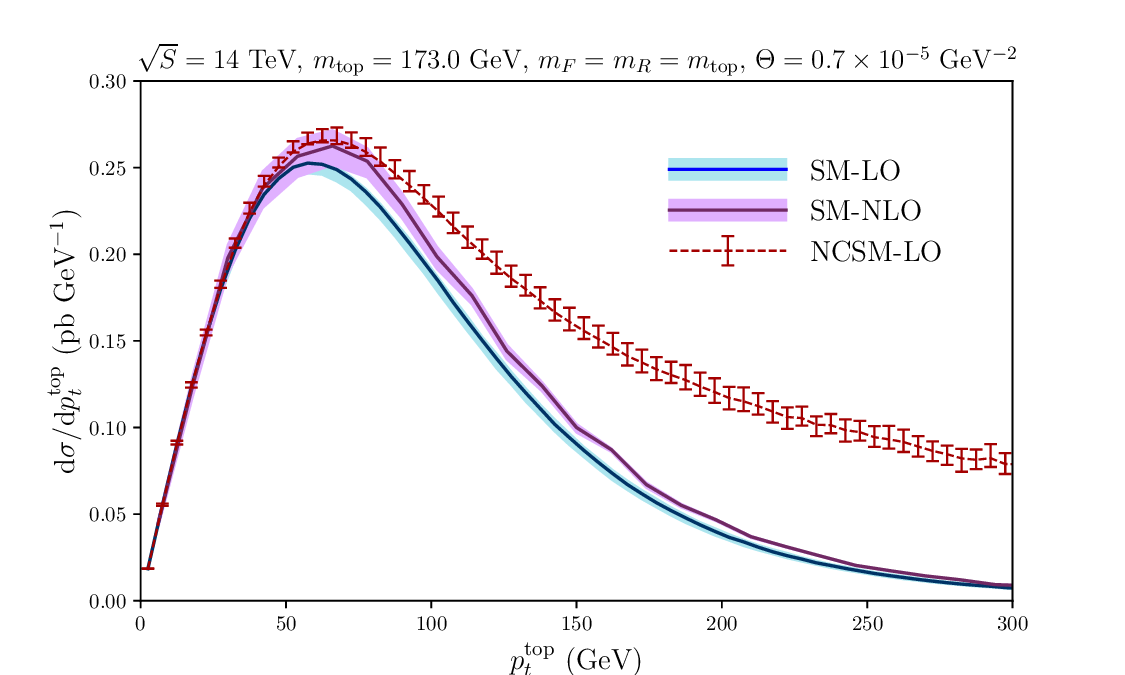}
		\vspace*{8pt}
		\caption{Top quark $p_t^{\rm{top}}$ distribution within single top production in $tW$ associated production. \protect\label{fig5}}
\end{figure}

We continue our discussion on the differential distributions for single top-quark production in the framework of NC geometry, focusing on the transverse momentum distribution of the top quark in single top-quark production associated with a $W$ boson. In Fig.~\ref{fig5}, we plot the transverse momentum distribution of the top quark in the SM at LO and NLO, as well as in the NCSM at LO. We observe that the behavior of the distributions is synchronized in the low $p_t^{\rm{top}}$ region, while the NC distribution deviates from the SM distribution as $p_t^{\rm{top}}$ increases, indicating the impact of NC corrections. This deviation becomes more pronounced with an increasing value of the NC parameter $\Theta$.
	
\subsubsection{Comparison with NNLO Predictions}

To obtain an accurate analysis of the effects of NC corrections on top-quark production with $X$ where ($X = \bar{t}, W^-$, and jet), we need to compare our results for the transverse momentum distribution with the NNLO distribution in the SM. In this section, we will perform this comparison. To achieve this, we use the results presented in Refs.~\cite{Kidonakis:2010dk, Berger:2017zof, Kidonakis:2021vob} for the transverse momentum distribution of top quarks at NNLO for top pair production, the $t$-channel of single top production, and the $tW$-channel of single top production, respectively.
	
\begin{figure}[ht]
		\centering
		\includegraphics[width=0.8\textwidth]{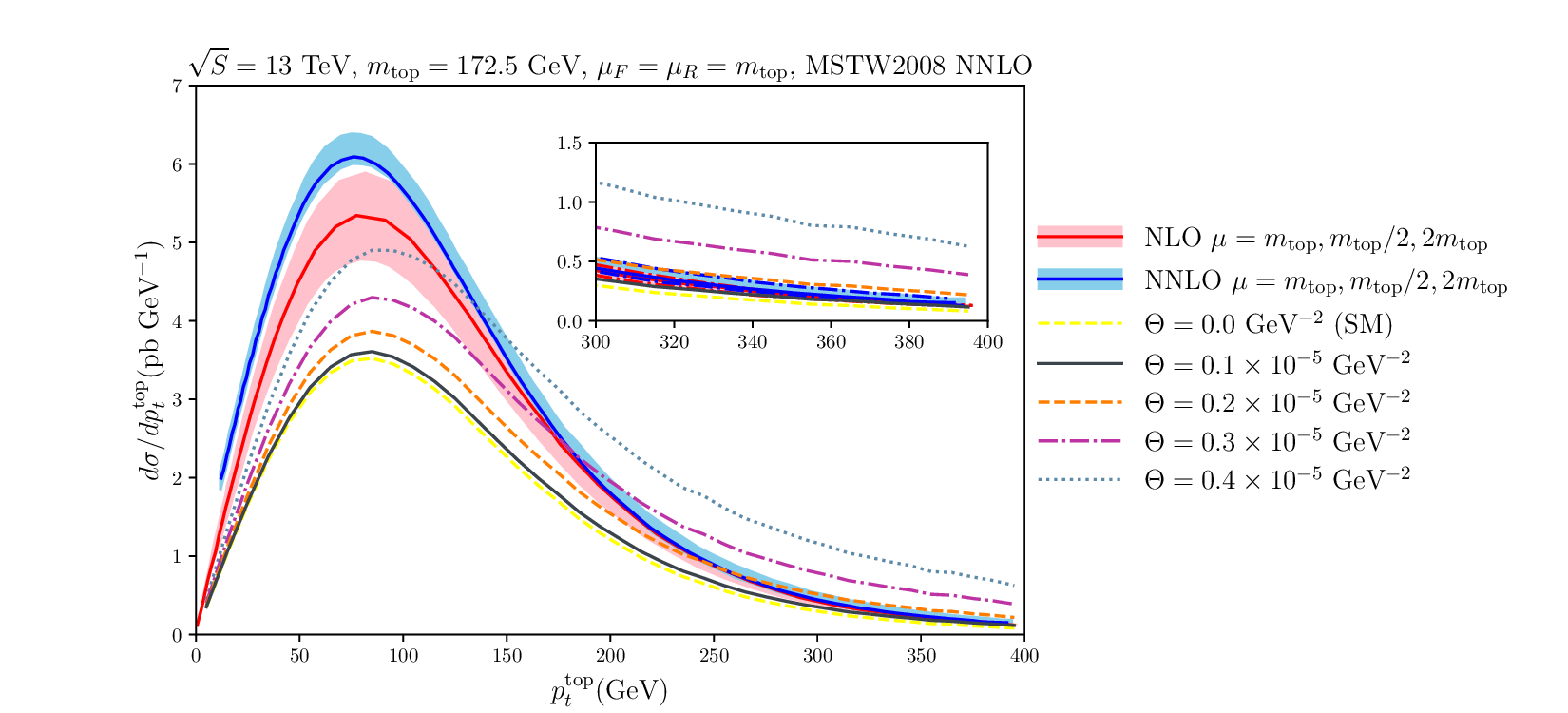}
		\vspace*{8pt}
		\caption{Comparison between top quark $p_t^{\rm{top}}$ distribution in pair top quark production within the NC-SM and fixed-order results up to NNLO. The fixed-order results are extracted from Ref.~\cite{Kidonakis:2010dk}. \protect\label{fig6}}
\end{figure}
	
In Fig.~\ref{fig6}, we display the results of top quark $p_t^{\rm{top}}$ distribution in pair top quark production at NNLO (from \cite{Kidonakis:2010dk}) and our results for top quark production within the NCSM, where we use the MSTW2008 NNLO \cite{martin2009parton} PDF set, and we fix the pole mass $m_{\mathrm{top}}$ of the top quark to the value $m_\mathrm{top} = 172.5$~GeV. 
	We observe that at the tail of the distribution, the fixed-order NNLO distribution curve is similar to the NLO curve, indicating that there are no significant effects of the NNLO corrections in this region. This means that we can still clearly distinguish the effects of NCG from NNLO corrections in this region. The NC corrections remain visible for values of $\Theta \gtrsim 0.3 \times 10^{-5} \, \mathrm{GeV}^{-2}$, which corresponds to a non-commutative energy scale $\Lambda_{\texttt{NC}} \sim 577.35 \, \mathrm{GeV}$.
\begin{figure}[ht]
		\centering
		\includegraphics[width=0.8\textwidth]{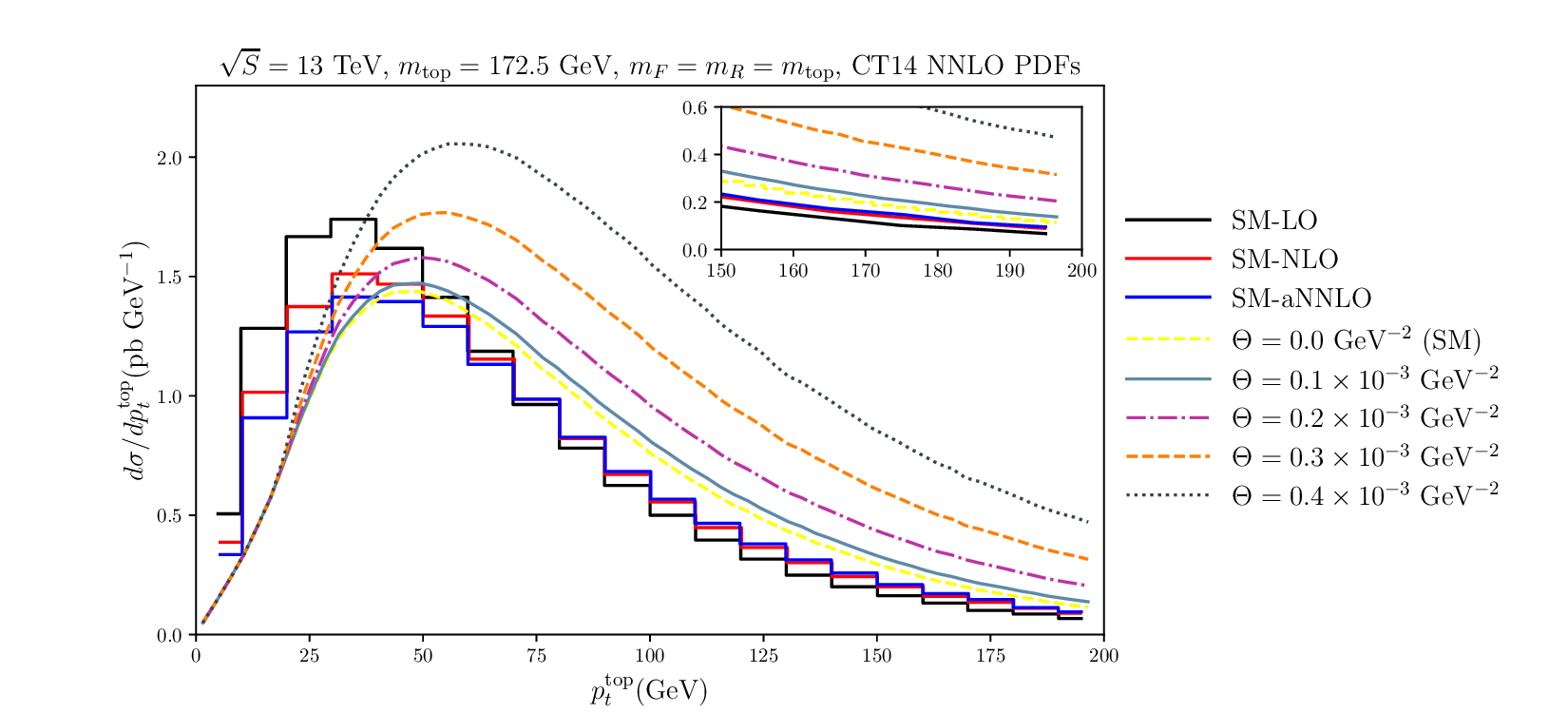}
		\vspace*{8pt}
		\caption{Comparison between top quark $p_t^{\rm{top}}$ distribution in single top quark production within the NC-SM and fixed-order results up to NNLO. The fixed-order results are extracted from Ref.~\cite{Berger:2017zof}. \protect\label{fig7}}
\end{figure}
	
Fig.~\ref{fig7} shows the results for the top quark $p_t^{\rm{top}}$ distribution in $t$-channel single top quark production at NNLO and our results for the top quark production within the NCSM. We observe that the NNLO and NLO curves converge, indicating that the higher-order corrections are insignificant. Consequently, we can still distinguish the effects of non-commutative geometry for values of $\Theta \gtrsim 0.2 \times 10^{-3} \, \mathrm{GeV}^{-2}$, which corresponds to $\Lambda_{\texttt{NC}} \sim 70.71 \, \mathrm{GeV}$.
\begin{figure}[ht]
		\centering
		\includegraphics[width=0.8\textwidth]{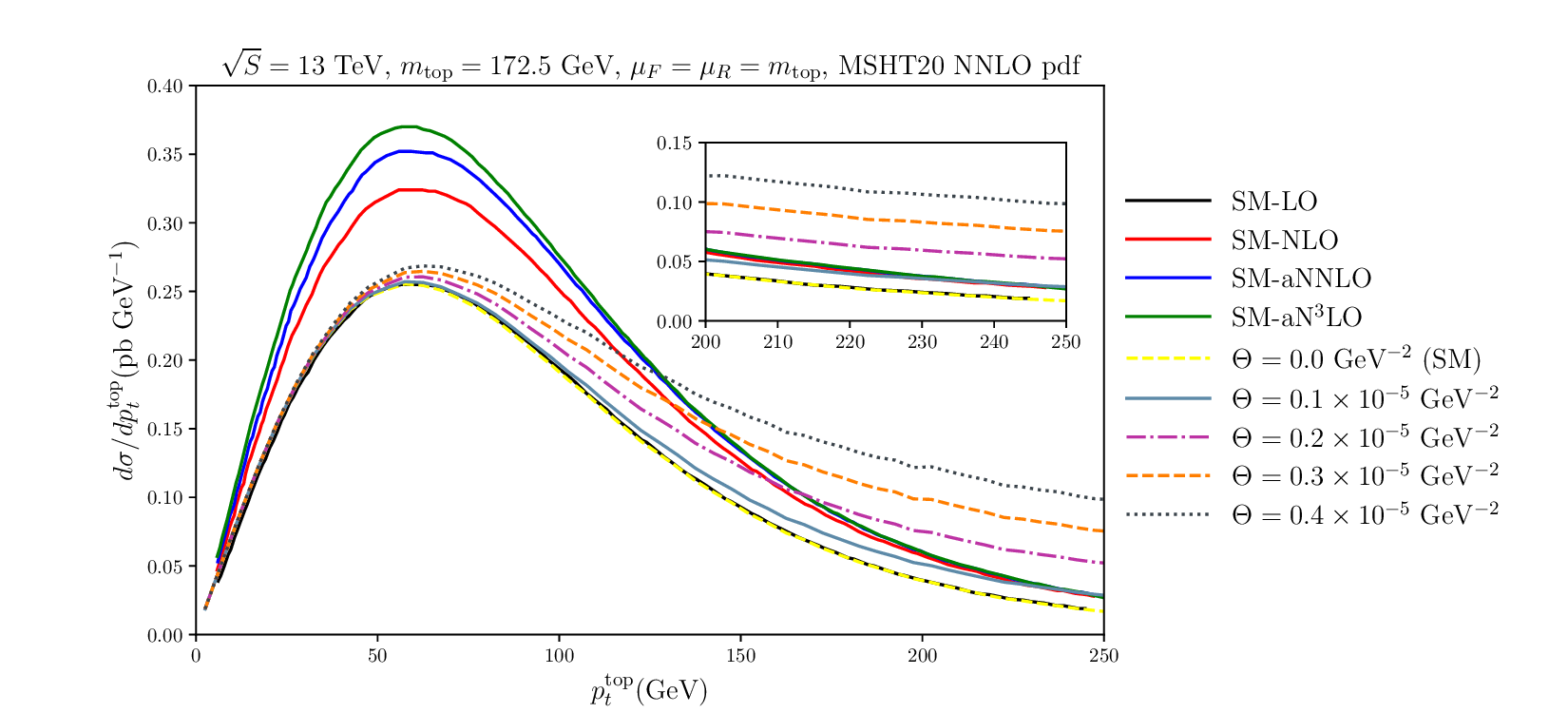}
		\vspace*{8pt}
		\caption{Comparison between top quark $p_t^{\rm{top}}$ distribution in $tW$ associated production within the NC-SM and fixed-order results up to NNLO. The fixed-order results are extracted from Ref.~\cite{Kidonakis:2021vob}. \protect\label{fig8}}
\end{figure}

Finally, Fig.~\ref{fig8} displays the top quark $p_t^{\rm{top}}$ distribution in $tW$ associated production at NNLO and our results within the NCSM. Similar to the previous figures, we observe that the NNLO and NLO curves converge at the tail of the distribution, indicating that the NNLO corrections have minimal impact. The effects of NCG remain distinguishable for $\Theta \gtrsim 0.2 \times 10^{-5} \, \mathrm{GeV}^{-2}$, corresponding to a non-commutative energy scale of $\Lambda_{\texttt{NC}} \sim 707.107 \, \mathrm{GeV}$.

\section{Conclusions}

In this work, we conducted a comprehensive study of top quark phenomenology within the framework of the NCSM. We derived the NC corrections to the squared amplitudes of the SM for both top quark pair production and single top quark production via the $t$-channel and $tW$-channel, noting that the $s$-channel remains unaffected by non-commutative geometry.
	
We proceeded to compute the total cross-sections at various center-of-mass energies for all processes, including top quark pair production and single top quark production within the NCSM, using \texttt{MadGraph5\_aMC@NLO}. This involved modifying event weights to incorporate non-commutative corrections to the SM amplitude squared. Additionally, we calculated the differential distributions in transverse momentum and rapidity of the top quark at leading order in the strong coupling constant. 
	
For single top quark production via the t-channel, we employed a simple matching formula that allowed us to extend the validity of the distribution to low \(p_t^{\rm{top}}\) values by showering (or resumming) the distribution within the SM and subsequently including non-commutative corrections at leading order. Finally, we compared the magnitude of non-commutative corrections for different values of \(\Theta\) with those obtained in QCD at NLO using the \texttt{MCFM} Monte Carlo program, and at NNLO using data from ref. \cite{Kidonakis:2010dk} for top quark pair production, \cite{Berger:2017zof} for the $t$-channel, and \cite{Kidonakis:2021vob} for the $tW$-channel of single top quark production.

\section*{Acknowledgments}

\begin{itemize}
\item This work is supported by the PRFU research project B00L02UN050120230003. The authors wish to thank the Algerian Ministry of Higher Education and Scientific Research and DGRSDT for their financial support.
\item Some of the numerical calculations presented here were performed on the HPC cluster at the University of Batna 2. We also gratefully acknowledge the LENREZA lab for their invaluable contributions and support.
\end{itemize}

\bibliographystyle{elsarticle-num} 
\bibliography{ref}

\begin{thebibliography}{10}
\expandafter\ifx\csname url\endcsname\relax
  \def\url#1{\texttt{#1}}\fi
\expandafter\ifx\csname urlprefix\endcsname\relax\def\urlprefix{URL }\fi
\expandafter\ifx\csname href\endcsname\relax
  \def\href#1#2{#2} \def\path#1{#1}\fi

\bibitem{Khachatryan_2015}
V.~Khachatryan, et~al., Measurement of the differential cross section for top
  quark pair production in pp collisions at $\sqrt{s} = 8$ tev, The European
  Physical Journal C 75~(11) (Nov 2015).
\newblock \href {https://doi.org/10.1140/epjc/s10052-015-3709-x}
  {\path{doi:10.1140/epjc/s10052-015-3709-x}}.

\bibitem{Khachatryan_2016}
V.~Khachatryan, et~al., Measurement of the $\mathrm{t}\overline{{\mathrm{t} }}$
  production cross section in the all-jets final state in pp collisions at
  $\sqrt{s}=8$ tev, The European Physical Journal C 76~(3) (Mar 2016).
\newblock \href {https://doi.org/10.1140/epjc/s10052-016-3956-5}
  {\path{doi:10.1140/epjc/s10052-016-3956-5}}.

\bibitem{Aad_2016a}
G.~Aad, et~al., Measurement of the differential cross-section of highly boosted
  top quarks as a function of their transverse momentum $\sqrt{s}=8$tev
  proton-proton collisions using the atlas detector, Physical Review D 93~(3)
  (Feb 2016).
\newblock \href {https://doi.org/10.1103/physrevd.93.032009}
  {\path{doi:10.1103/physrevd.93.032009}}.

\bibitem{Aad_2016b}
G.~Aad, et~al., Measurements of top-quark pair differential cross-sections in
  the lepton+jets channel in pp collisions at $\sqrt{s}=8\,~{\mathrm {tev} }$
  using the atlas detector, The European Physical Journal C 76~(10) (Oct 2016).
\newblock \href {https://doi.org/10.1140/epjc/s10052-016-4366-4}
  {\path{doi:10.1140/epjc/s10052-016-4366-4}}.

\bibitem{NASON1988607}
P.~Nason, S.~Dawson, R.~Ellis, The total cross section for the production of
  heavy quarks in hadronic collisions, Nuclear Physics B 303~(4) (1988)
  607--633.
\newblock \href {https://doi.org/https://doi.org/10.1016/0550-3213(88)90422-1}
  {\path{doi:https://doi.org/10.1016/0550-3213(88)90422-1}}.

\bibitem{BEENAKKER1991507}
W.~Beenakker, W.~{Van Neerven}, R.~Meng, G.~Schuler, J.~Smith, Qcd corrections
  to heavy quark production in hadron-hadron collisions, Nuclear Physics B
  351~(3) (1991) 507--560.
\newblock \href {https://doi.org/https://doi.org/10.1016/S0550-3213(05)80032-X}
  {\path{doi:https://doi.org/10.1016/S0550-3213(05)80032-X}}.

\bibitem{MANGANO1992295}
M.~L. Mangano, P.~Nason, G.~Ridolfi, Heavy-quark correlations in hadron
  collisions at next-to-leading order, Nuclear Physics B 373~(2) (1992)
  295--345.
\newblock \href {https://doi.org/https://doi.org/10.1016/0550-3213(92)90435-E}
  {\path{doi:https://doi.org/10.1016/0550-3213(92)90435-E}}.

\bibitem{PhysRevD.40.54}
W.~Beenakker, H.~Kuijf, W.~L. van Neerven, J.~Smith, Qcd corrections to
  heavy-quark production in $p\overline{p}$ collisions, Phys. Rev. D 40 (1989)
  54--82.
\newblock \href {https://doi.org/https://doi.org/10.1103/PhysRevD.40.54}
  {\path{doi:https://doi.org/10.1103/PhysRevD.40.54}}.

\bibitem{beenakker1994electroweak}
W.~Beenakker, A.~Denner, W.~Hollik, R.~Mertig, T.~Sack, D.~Wackeroth,
  Electroweak one-loop contributions to top pair production in hadron
  colliders, Nuclear Physics B 411~(2-3) (1994) 343--380.
\newblock \href {https://doi.org/https://doi.org/10.1016/0550-3213(94)90454}
  {\path{doi:https://doi.org/10.1016/0550-3213(94)90454}}.

\bibitem{kuhn2007weak}
J.~H. Kuhn, A.~Scharf, P.~Uwer, {Electroweak effects in top-quark pair
  production at hadron colliders}, Eur. Phys. J. C 51 (2007) 37--53.
\newblock \href {http://arxiv.org/abs/hep-ph/0610335}
  {\path{arXiv:hep-ph/0610335}}, \href
  {https://doi.org/10.1140/epjc/s10052-007-0275-x}
  {\path{doi:10.1140/epjc/s10052-007-0275-x}}.

\bibitem{abazov2009observation}
D.~collaboration, Observation of single top-quark production, Physical review
  letters 103~(9) (2009) 092001.
\newblock \href
  {https://doi.org/https://doi.org/10.1103/PhysRevLett.103.092001}
  {\path{doi:https://doi.org/10.1103/PhysRevLett.103.092001}}.

\bibitem{CDF:2009itk}
C.~collaboration, First observation of electroweak single top quark production,
  Phys. Rev. Lett. 103 (2009).
\newblock \href {https://doi.org/10.1103/PhysRevLett.103.092002}
  {\path{doi:10.1103/PhysRevLett.103.092002}}.

\bibitem{campbell2021single}
J.~Campbell, T.~Neumann, Z.~Sullivan, Single-top-quark production in the
  t-channel at nnlo, Journal of High Energy Physics 2021~(2) (2021) 1--41.
\newblock \href {https://doi.org/https://doi.org/10.1007/JHEP02(2021)040}
  {\path{doi:https://doi.org/10.1007/JHEP02(2021)040}}.

\bibitem{ott2012search}
J.~Ott, Search for single top tw associated production in the dilepton channel
  at cms, in: EPJ Web of Conferences, Vol.~28, EDP Sciences, 2012, p. 12041.
\newblock \href {https://doi.org/https://doi.org/10.1051/epjconf/20122812041}
  {\path{doi:https://doi.org/10.1051/epjconf/20122812041}}.

\bibitem{CMS:2017mpr}
A.~M. Sirunyan, et~al., {Measurement of the top quark mass using single top
  quark events in proton-proton collisions at $\sqrt{s}= 8$ TeV}, Eur. Phys. J.
  C 77~(5) (2017) 354.
\newblock \href {http://arxiv.org/abs/1703.02530} {\path{arXiv:1703.02530}},
  \href {https://doi.org/10.1140/epjc/s10052-017-4912-8}
  {\path{doi:10.1140/epjc/s10052-017-4912-8}}.

\bibitem{khachatryan2016measurement}
V.~Khachatryan, A.~M. Sirunyan, A.~Tumasyan, W.~Adam, E.~Asilar, T.~Bergauer,
  J.~Brandstetter, E.~Brondolin, M.~Dragicevic, J.~Er{\"o}, et~al., Measurement
  of top quark polarisation in t-channel single top quark production, Journal
  of High Energy Physics 2016~(4) (2016) 1--42.
\newblock \href {https://doi.org/https://doi.org/10.1007/JHEP04(2016)073}
  {\path{doi:https://doi.org/10.1007/JHEP04(2016)073}}.

\bibitem{aaboud2017fiducial}
M.~Aaboud, G.~Aad, B.~Abbott, J.~Abdallah, O.~Abdinov, B.~Abeloos, O.~AbouZeid,
  N.~Abraham, H.~Abramowicz, H.~Abreu, et~al., Fiducial, total and differential
  cross-section measurements of t-channel single top-quark production in pp
  collisions at 8 tev using data collected by the atlas detector, The European
  Physical Journal C 77 (2017) 1--46.
\newblock \href
  {https://doi.org/https://doi.org/10.1140/epjc/s10052-017-5061-9}
  {\path{doi:https://doi.org/10.1140/epjc/s10052-017-5061-9}}.

\bibitem{sirunyan2020measurement}
A.~M. Sirunyan, A.~Tumasyan, W.~Adam, F.~Ambrogi, T.~Bergauer, J.~Brandstetter,
  M.~Dragicevic, J.~Er{\"o}, A.~E. Del~Valle, M.~Flechl, et~al., Measurement of
  differential cross sections and charge ratios for t-channel single top quark
  production in proton--proton collisions at $\sqrt{s}$= 13 tev, The European
  Physical Journal C 80~(5) (2020) 1--37.
\newblock \href
  {https://doi.org/https://doi.org/10.1140/epjc/s10052-020-7858-1}
  {\path{doi:https://doi.org/10.1140/epjc/s10052-020-7858-1}}.

\bibitem{aad2016search}
G.~Aad, B.~Abbott, J.~Abdallah, R.~Aben, M.~Abolins, O.~AbouZeid,
  H.~Abramowicz, H.~Abreu, R.~Abreu, Y.~Abulaiti, et~al., Search for anomalous
  couplings in the w tb vertex from the measurement of double differential
  angular decay rates of single top quarks produced in the t-channel with the
  atlas detector, Journal of high energy physics 2016~(4) (2016) 1--46.
\newblock \href {https://doi.org/https://doi.org/10.1007/JHEP04(2016)023}
  {\path{doi:https://doi.org/10.1007/JHEP04(2016)023}}.

\bibitem{CMS:2016uzc}
V.~Khachatryan, et~al., {Search for anomalous Wtb couplings and
  flavour-changing neutral currents in t-channel single top quark production in
  pp collisions at $\sqrt{s} =$ 7 and 8 TeV}, JHEP 02 (2017) 028.
\newblock \href {http://arxiv.org/abs/1610.03545} {\path{arXiv:1610.03545}},
  \href {https://doi.org/10.1007/JHEP02(2017)028}
  {\path{doi:10.1007/JHEP02(2017)028}}.

\bibitem{ATLAS:2017ygi}
M.~Aaboud, et~al., {Probing the W tb vertex structure in t-channel
  single-top-quark production and decay in pp collisions at $ \sqrt{s}=8 $ TeV
  with the ATLAS detector}, JHEP 04 (2017) 124.
\newblock \href {http://arxiv.org/abs/1702.08309} {\path{arXiv:1702.08309}},
  \href {https://doi.org/10.1007/JHEP04(2017)124}
  {\path{doi:10.1007/JHEP04(2017)124}}.

\bibitem{CDF:2015gsg}
T.~A. Aaltonen, et~al., {Tevatron Combination of Single-Top-Quark Cross
  Sections and Determination of the Magnitude of the Cabibbo-Kobayashi-Maskawa
  Matrix Element $\bf V_{tb}$}, Phys. Rev. Lett. 115~(15) (2015) 152003.
\newblock \href {http://arxiv.org/abs/1503.05027} {\path{arXiv:1503.05027}},
  \href {https://doi.org/10.1103/PhysRevLett.115.152003}
  {\path{doi:10.1103/PhysRevLett.115.152003}}.

\bibitem{chatrchyan2011measurement}
S.~Chatrchyan, V.~Khachatryan, A.~M. Sirunyan, A.~Tumasyan, W.~Adam,
  T.~Bergauer, M.~Dragicevic, J.~Er{\"o}, C.~Fabjan, M.~Friedl, et~al.,
  Measurement of the t-channel single top quark production cross section in p p
  collisions at s= 7 tev, Physical Review Letters 107~(9) (2011) 091802.
\newblock \href {https://doi.org/10.1103/PhysRevLett.107.091802}
  {\path{doi:10.1103/PhysRevLett.107.091802}}.

\bibitem{CMS:2014mgj}
V.~Khachatryan, et~al., {Measurement of the t-channel single-top-quark
  production cross section and of the $\mid V_{tb} \mid$ CKM matrix element in
  pp collisions at $\sqrt{s}$= 8 TeV}, JHEP 06 (2014) 090.
\newblock \href {http://arxiv.org/abs/1403.7366} {\path{arXiv:1403.7366}},
  \href {https://doi.org/10.1007/JHEP06(2014)090}
  {\path{doi:10.1007/JHEP06(2014)090}}.

\bibitem{daubie2013measurement}
E.~Daubie, N.~Beliy, T.~Caebergs, G.~Hammad, C.~Collaboration, et~al.,
  Measurement of differential top-quark-pair production cross sections in pp
  collisions at sqrt (s)= 7 tev, European Physical Journal C. Particles and
  Fields 73 (2013).
\newblock \href
  {https://doi.org/https://doi.org/10.1140/epjc/s10052-013-2339-4}
  {\path{doi:https://doi.org/10.1140/epjc/s10052-013-2339-4}}.

\bibitem{ATLAS:2015dbj}
G.~Aad, et~al., {Differential top-antitop cross-section measurements as a
  function of observables constructed from final-state particles using pp
  collisions at $\sqrt{s}=7$ TeV in the ATLAS detector}, JHEP 06 (2015) 100.
\newblock \href {http://arxiv.org/abs/1502.05923} {\path{arXiv:1502.05923}},
  \href {https://doi.org/10.1007/JHEP06(2015)100}
  {\path{doi:10.1007/JHEP06(2015)100}}.

\bibitem{douglas2001noncommutative}
M.~R. Douglas, N.~A. Nekrasov, {Noncommutative field theory}, Rev. Mod. Phys.
  73 (2001) 977--1029.
\newblock \href {http://arxiv.org/abs/hep-th/0106048}
  {\path{arXiv:hep-th/0106048}}, \href
  {https://doi.org/10.1103/RevModPhys.73.977}
  {\path{doi:10.1103/RevModPhys.73.977}}.

\bibitem{madore2000gauge}
J.~Madore, S.~Schraml, P.~Schupp, J.~Wess, Gauge theory on noncommutative
  spaces, The European Physical Journal C-Particles and Fields 16 (2000)
  161--167.
\newblock \href {https://doi.org/https://doi.org/10.1007/s100520050012}
  {\path{doi:https://doi.org/10.1007/s100520050012}}.

\bibitem{jurco2000enveloping}
B.~Jurco, S.~Schraml, P.~Schupp, J.~Wess, Enveloping algebra-valued gauge
  transformations for non-abelian gauge groups on non-commutative spaces, The
  European Physical Journal C-Particles and Fields 17 (2000) 521--526.
\newblock \href {https://doi.org/https://doi.org/10.1007/s100520000487}
  {\path{doi:https://doi.org/10.1007/s100520000487}}.

\bibitem{jurvco2001nonabelian}
B.~Jur{\v{c}}o, P.~Schupp, J.~Wess, Nonabelian noncommutative gauge theory via
  noncommutative extra dimensions, Nuclear Physics B 604~(1-2) (2001) 148--180.
\newblock \href {https://doi.org/https://doi.org/10.1016/S0550-3213(01)00191-2}
  {\path{doi:https://doi.org/10.1016/S0550-3213(01)00191-2}}.

\bibitem{Calmet:2001na}
X.~Calmet, B.~Jurco, P.~Schupp, J.~Wess, M.~Wohlgenannt, {The Standard model on
  noncommutative space-time}, Eur. Phys. J. C 23 (2002) 363--376.
\newblock \href {http://arxiv.org/abs/hep-ph/0111115}
  {\path{arXiv:hep-ph/0111115}}, \href {https://doi.org/10.1007/s100520100873}
  {\path{doi:10.1007/s100520100873}}.

\bibitem{Mahajan:2003ze}
N.~Mahajan, {$t \to b W$ in noncommutative standard model}, Phys. Rev. D 68
  (2003) 095001.
\newblock \href {http://arxiv.org/abs/hep-ph/0304235}
  {\path{arXiv:hep-ph/0304235}}, \href
  {https://doi.org/10.1103/PhysRevD.68.095001}
  {\path{doi:10.1103/PhysRevD.68.095001}}.

\bibitem{MohammadiNajafabadi:2008zlg}
M.~Mohammadi~Najafabadi, {Noncommutative Standard Model in Top Quark Sector},
  Phys. Rev. D 77 (2008) 116011.
\newblock \href {http://arxiv.org/abs/0803.2340} {\path{arXiv:0803.2340}},
  \href {https://doi.org/10.1103/PhysRevD.77.116011}
  {\path{doi:10.1103/PhysRevD.77.116011}}.

\bibitem{MohammadiNajafabadi:2006iu}
M.~Mohammadi~Najafabadi, {Semi-leptonic decay of a polarized top quark in the
  noncommutative standard model}, Phys. Rev. D 74 (2006) 025021.
\newblock \href {http://arxiv.org/abs/hep-ph/0606017}
  {\path{arXiv:hep-ph/0606017}}, \href
  {https://doi.org/10.1103/PhysRevD.74.025021}
  {\path{doi:10.1103/PhysRevD.74.025021}}.

\bibitem{Fisli:2020vzt}
M.~Fisli, N.~Mebarki, {Top Quark Pair-Production in Noncommutative Standard
  Model}, Adv. High Energy Phys. 2020 (2020) 7279627.
\newblock \href {https://doi.org/10.1155/2020/7279627}
  {\path{doi:10.1155/2020/7279627}}.

\bibitem{Rezaei:2018mlk}
Z.~Rezaei, R.~Salehi, {Azimuthal correlation function of polarized top quark in
  noncommutative space\textendash{}time}, Annals Phys. 406 (2019) 71--85.
\newblock \href {http://arxiv.org/abs/1804.07688} {\path{arXiv:1804.07688}},
  \href {https://doi.org/10.1016/j.aop.2019.04.004}
  {\path{doi:10.1016/j.aop.2019.04.004}}.

\bibitem{Filk:1996dm}
T.~Filk, {Divergencies in a field theory on quantum space}, Phys. Lett. B 376
  (1996) 53--58.
\newblock \href {https://doi.org/10.1016/0370-2693(96)00024-X}
  {\path{doi:10.1016/0370-2693(96)00024-X}}.

\bibitem{riad2000noncommutative}
I.~F. Riad, M.~M. Sheikh-Jabbari, {Noncommutative QED and anomalous dipole
  moments}, JHEP 08 (2000) 045.
\newblock \href {http://arxiv.org/abs/hep-th/0008132}
  {\path{arXiv:hep-th/0008132}}, \href
  {https://doi.org/10.1088/1126-6708/2000/08/045}
  {\path{doi:10.1088/1126-6708/2000/08/045}}.

\bibitem{seiberg1999string}
N.~Seiberg, E.~Witten, String theory and noncommutative geometry, Journal of
  High Energy Physics 1999~(09) (1999) 032.
\newblock \href {https://doi.org/10.1088/1126-6708/1999/09/032}
  {\path{doi:10.1088/1126-6708/1999/09/032}}.

\bibitem{melic2005standard}
B.~Melic, K.~Passek-Kumericki, J.~Trampetic, P.~Schupp, M.~Wohlgenannt, {The
  Standard model on non-commutative space-time: Electroweak currents and Higgs
  sector}, Eur. Phys. J. C 42 (2005) 483--497.
\newblock \href {http://arxiv.org/abs/hep-ph/0502249}
  {\path{arXiv:hep-ph/0502249}}, \href
  {https://doi.org/10.1140/epjc/s2005-02318-6}
  {\path{doi:10.1140/epjc/s2005-02318-6}}.

\bibitem{uelker2008seiberg}
K.~Ulker, B.~Yapiskan, {Seiberg-Witten maps to all orders}, Phys. Rev. D 77
  (2008) 065006.
\newblock \href {http://arxiv.org/abs/0712.0506} {\path{arXiv:0712.0506}},
  \href {https://doi.org/10.1103/PhysRevD.77.065006}
  {\path{doi:10.1103/PhysRevD.77.065006}}.

\bibitem{ulker2012all}
K.~Ulker, {On the All Order Solutions of Seiberg-Witten Map for Noncommutative
  Gauge Theories}, Int. J. Mod. Phys. Conf. Ser. 13 (2012) 191--198.
\newblock \href {http://arxiv.org/abs/1201.2192} {\path{arXiv:1201.2192}},
  \href {https://doi.org/10.1142/S201019451200685X}
  {\path{doi:10.1142/S201019451200685X}}.

\bibitem{Jurco:2001rq}
B.~Jurco, L.~Moller, S.~Schraml, P.~Schupp, J.~Wess, {Construction of
  nonAbelian gauge theories on noncommutative spaces}, Eur. Phys. J. C 21
  (2001) 383--388.
\newblock \href {http://arxiv.org/abs/hep-th/0104153}
  {\path{arXiv:hep-th/0104153}}, \href {https://doi.org/10.1007/s100520100731}
  {\path{doi:10.1007/s100520100731}}.

\bibitem{Maltoni:2002qb}
F.~Maltoni, T.~Stelzer, {MadEvent: Automatic event generation with MadGraph},
  JHEP 02 (2003) 027.
\newblock \href {https://doi.org/10.1088/1126-6708/2003/02/027}
  {\path{doi:10.1088/1126-6708/2003/02/027}}.

\bibitem{Alwall:2014hca}
J.~Alwall, R.~Frederix, S.~Frixione, V.~Hirschi, F.~Maltoni, O.~Mattelaer,
  H.~S. Shao, T.~Stelzer, P.~Torrielli, M.~Zaro, {The automated computation of
  tree-level and next-to-leading order differential cross sections, and their
  matching to parton shower simulations}, JHEP 07 (2014) 079.
\newblock \href {https://doi.org/10.1007/JHEP07(2014)079}
  {\path{doi:10.1007/JHEP07(2014)079}}.

\bibitem{Conte:2012fm}
E.~Conte, B.~Fuks, G.~Serret, {MadAnalysis 5, A User-Friendly Framework for
  Collider Phenomenology}, Comput. Phys. Commun. 184 (2013) 222--256.
\newblock \href {https://doi.org/10.1016/j.cpc.2012.09.009}
  {\path{doi:10.1016/j.cpc.2012.09.009}}.

\bibitem{gomis2000space}
J.~Gomis, T.~Mehen, {Space-time noncommutative field theories and unitarity},
  Nucl. Phys. B 591 (2000) 265--276.
\newblock \href {http://arxiv.org/abs/hep-th/0005129}
  {\path{arXiv:hep-th/0005129}}, \href
  {https://doi.org/10.1016/S0550-3213(00)00525-3}
  {\path{doi:10.1016/S0550-3213(00)00525-3}}.

\bibitem{ParticleDataGroup:2022pth}
R.~L. Workman, et~al., {Review of Particle Physics}, PTEP 2022 (2022) 083C01.
\newblock \href {https://doi.org/10.1093/ptep/ptac097}
  {\path{doi:10.1093/ptep/ptac097}}.

\bibitem{kidonakis2010two}
N.~Kidonakis, {Two-loop soft anomalous dimensions for single top quark
  associated production with a $W^-$ or $H^-$}, Phys. Rev. D 82 (2010) 054018.
\newblock \href {http://arxiv.org/abs/1005.4451} {\path{arXiv:1005.4451}},
  \href {https://doi.org/10.1103/PhysRevD.82.054018}
  {\path{doi:10.1103/PhysRevD.82.054018}}.

\bibitem{sirunyan2017cross}
A.~M. Sirunyan, A.~Tumasyan, W.~Adam, E.~Asilar, T.~Bergauer, J.~Brandstetter,
  E.~Brondolin, M.~Dragicevic, J.~Er{\"o}, M.~Flechl, et~al., Cross section
  measurement of t-channel single top quark production in pp collisions at s=
  13tev, Physics Letters B 772 (2017) 752--776.

\bibitem{harris2002fully}
B.~W. Harris, E.~Laenen, L.~Phaf, Z.~Sullivan, S.~Weinzierl, {The Fully
  Differential Single Top Quark Cross-Section in Next to Leading Order QCD},
  Phys. Rev. D 66 (2002) 054024.
\newblock \href {http://arxiv.org/abs/hep-ph/0207055}
  {\path{arXiv:hep-ph/0207055}}, \href
  {https://doi.org/10.1103/PhysRevD.66.054024}
  {\path{doi:10.1103/PhysRevD.66.054024}}.

\bibitem{ball2017parton}
R.~D. Ball, et~al., {Parton distributions from high-precision collider data},
  Eur. Phys. J. C 77~(10) (2017) 663.
\newblock \href {http://arxiv.org/abs/1706.00428} {\path{arXiv:1706.00428}},
  \href {https://doi.org/10.1140/epjc/s10052-017-5199-5}
  {\path{doi:10.1140/epjc/s10052-017-5199-5}}.

\bibitem{buckley2015lhapdf6}
A.~Buckley, J.~Ferrando, S.~Lloyd, K.~Nordstr{\"o}m, B.~Page, M.~R{\"u}fenacht,
  M.~Sch{\"o}nherr, G.~Watt, Lhapdf6: parton density access in the lhc
  precision era, The European Physical Journal C 75 (2015) 1--20.
\newblock \href
  {https://doi.org/https://doi.org/10.1140/epjc/s10052-015-3318-8}
  {\path{doi:https://doi.org/10.1140/epjc/s10052-015-3318-8}}.

\bibitem{Campbell:2019dru}
J.~Campbell, T.~Neumann, {Precision Phenomenology with MCFM}, JHEP 12 (2019)
  034.
\newblock \href {https://doi.org/10.1007/JHEP12(2019)034}
  {\path{doi:10.1007/JHEP12(2019)034}}.

\bibitem{Kidonakis:2010dk}
N.~Kidonakis, {Next-to-next-to-leading soft-gluon corrections for the top quark
  cross section and transverse momentum distribution}, Phys. Rev. D 82 (2010)
  114030.
\newblock \href {http://arxiv.org/abs/1009.4935} {\path{arXiv:1009.4935}},
  \href {https://doi.org/10.1103/PhysRevD.82.114030}
  {\path{doi:10.1103/PhysRevD.82.114030}}.

\bibitem{Berger:2017zof}
E.~L. Berger, J.~Gao, H.~X. Zhu, {Differential Distributions for t-channel
  Single Top-Quark Production and Decay at Next-to-Next-to-Leading Order in
  QCD}, JHEP 11 (2017) 158.
\newblock \href {http://arxiv.org/abs/1708.09405} {\path{arXiv:1708.09405}},
  \href {https://doi.org/10.1007/JHEP11(2017)158}
  {\path{doi:10.1007/JHEP11(2017)158}}.

\bibitem{Kidonakis:2021vob}
N.~Kidonakis, N.~Yamanaka, {Higher-order corrections for $tW$ production at
  high-energy hadron colliders}, JHEP 05 (2021) 278.
\newblock \href {http://arxiv.org/abs/2102.11300} {\path{arXiv:2102.11300}},
  \href {https://doi.org/10.1007/JHEP05(2021)278}
  {\path{doi:10.1007/JHEP05(2021)278}}.

\bibitem{martin2009parton}
A.~D. Martin, W.~J. Stirling, R.~S. Thorne, G.~Watt, Parton distributions for
  the lhc, The European Physical Journal C 63~(2) (2009) 189--285.
\newblock \href
  {https://doi.org/https://doi.org/10.1140/epjc/s10052-009-1072-5}
  {\path{doi:https://doi.org/10.1140/epjc/s10052-009-1072-5}}.

\end{thebibliography}

\end{document}